\documentclass{SciPost}

\hypersetup{
    colorlinks,
    linkcolor={red!50!black},
    citecolor={blue!50!black},
    urlcolor={blue!80!black}
}

\usepackage[bitstream-charter]{mathdesign}
\DeclareSymbolFont{usualmathcal}{OMS}{cmsy}{m}{n}
\DeclareSymbolFontAlphabet{\mathcal}{usualmathcal}

\newcommand{\Z}{\mathbb{Z}}

\newcommand{\R}{\mathbb{R}}
\newcommand{\C}{\mathbb{C}}

\newcommand{\norm}[1]{\left\lVert#1\right\rVert}

\newtheorem{definition}{Definition}[section]

\newtheorem{remark}[definition]{Remark}

\newtheorem{theorem}[definition]{Theorem}

\newtheorem{proposition}[definition]{Proposition}

\usepackage{bbm}

\fancypagestyle{SPstyle}{
\fancyhf{}
\lhead{\colorbox{scipostblue}{\bf \color{white} ~SciPost Physics }}
\rhead{{\bf \color{scipostdeepblue} ~Submission }}

\fancyfoot[C]{\textbf{\thepage}}
}

\begin{document}

\pagestyle{SPstyle}

\begin{center}{\Large \textbf{\color{scipostdeepblue}{
Global Gauge Symmetries and Spatial Asymptotic Boundary Conditions in Yang-Mills Theory\\
}}}\end{center}

\begin{center}\textbf{
Silvester G.A. Borsboom\textsuperscript{1$\star$} and
Hessel B. Posthuma\textsuperscript{2$\dagger$}
}\end{center}

\begin{center}
{\bf 1} Institute for Mathematics, Astrophysics and Particle Physics, Radboud University
\\
{\bf 2} Korteweg-de Vries Institute for Mathematics, University of Amsterdam
\\[\baselineskip]
$\star$ \href{mailto:email1}{\small silvester.borsboom@ru.nl}\,,\quad
$\dagger$ \href{mailto:email2}{\small h.b.posthuma@uva.nl}
\end{center}

\section*{\color{scipostdeepblue}{Abstract}}
\textbf{\boldmath{%
In Yang-Mills theory on a Euclidean Cauchy surface, the physical gauge group is often taken to be $\mathcal{G}^I/\mathcal{G}^\infty_0$, where $\mathcal{G}^I$ consists of boundary-preserving gauge transformations asymptoting to a constant, and $\mathcal{G}^\infty_0$ consists of transformations generated by the Gauss law constraint. We rigorously derive this physical gauge group for both Abelian and non-Abelian theories. A key result is that restricting to $\mathcal{G}^I$ follows from the structure of the instantaneous state space on which the instantaneous Lagrangian is defined. We extend our analysis to Yang-Mills-Higgs theory, showing that boundary conditions and the physical gauge group differ between the unbroken and broken phases.\footnote{This article grew out of the master thesis \cite{borsboom2024spontaneous} of the corresponding author SB, supervised by the other author and by Sebastian de Haro.}
}}

\vspace{\baselineskip}

\noindent\textcolor{white!90!black}{%
\fbox{\parbox{0.975\linewidth}{%
\textcolor{white!40!black}{\begin{tabular}{lr}%
  \begin{minipage}{0.6\textwidth}%
    {\small Copyright attribution to authors. \newline
    This work is a submission to SciPost Physics. \newline
    License information to appear upon publication. \newline
    Publication information to appear upon publication.}
  \end{minipage} & \begin{minipage}{0.4\textwidth}
    {\small Received Date \newline Accepted Date \newline Published Date}%
  \end{minipage}
\end{tabular}}
}}
}


\vspace{10pt}
\noindent\rule{\textwidth}{1pt}
\tableofcontents
\noindent\rule{\textwidth}{1pt}
\vspace{10pt}


\section{Introduction}\label{intro}

The physical status of gauge symmetries is a central topic in contemporary physics, both in Yang-Mills theory and general relativity. The term ``gauge'' is sometimes used as a synonym for ``unphysical'' or ``empirically insignificant,'' but gauge transformations can acquire a physical meaning in the presence of boundaries \cite{SNIATYCKI1988291,sniatycskiboundarycondspatboundeddom,gomesUnifiedGeometricFramework2018,gomesUnifiedGeometricFramework2019,gomesGaugingBoundaryFieldspace2019,gomesQuasilocalDegreesFreedom2021,rielloEdgeModesEdge2021,rielloHamiltonianGaugeTheory2024,rielloNullHamiltonianYang2025}. A well-known empirical example is the Josephson current flowing between two superconductors whose boundaries are brought close together \cite{josephsonPossibleNewEffects1962}. This current depends only on the relative difference between the \textit{global} $U(1)$ phases of the superconductors' Ginzburg-Landau order parameters, suggesting that global gauge symmetries are physical. Similarly, some gauge symmetries are physical on \textit{asymptotic} boundaries. For instance, the asymptotic symmetry group of gravity at null infinity in asymptotically flat spacetimes is the well-known BMS group \cite{bondiGravitationalWavesGeneral1962,sachsAsymptoticSymmetriesGravitational1962,henneauxAsymptoticStructureGravity2020}, and asymptotic symmetries of Yang-Mills fields on both the null and spatial conformal boundaries of Minkowski spacetime are studied in the context of celestial holography, see e.g. \cite{stromingerAsymptoticSymmetriesYangMills2014,stromingerLecturesInfraredStructure2018,henneauxAsymptoticSymmetriesElectromagnetism2018,pasterskiCelestialHolography2021,Tanzi:2020fmt,rielloNullHamiltonianYang2025}. The general idea is that the asymptotic symmetry group consists of all ``allowed symmetries'' quotiented by all ``trivial symmetries'' \cite{stromingerLecturesInfraredStructure2018}. Here ``allowed'' means those symmetries that respect the boundary conditions of the system and ``trivial'' means those symmetries that have no physical effect on the system. Following Dirac, in this article we identify the trivial symmetries of a gauge theory as those transformations that are generated by the Hamiltonian constraints of the theory.

Our aim is to rigorously derive the quotient of boundary-preserving gauge symmetries by trivial gauge symmetries for the specific case of Yang-Mills and Yang-Mills-Higgs theories on a Cauchy surface $\Sigma$ isomorphic to $\R^3$. Our motivation to do so comes from the desire to understand the physical content of the Higgs mechanism \cite{higgsBrokenSymmetriesMasses1964,higgsBrokenSymmetriesMasses1964a,englertBrokenSymmetryMass1964,guralnikGlobalConservationLaws1964}, which is thought to have happened at a particular instant in time during the electroweak phase transition. For this reason we introduce a 3+1 split $\Sigma\times \R$ of spacetime and discuss \textit{instantaneous} spatial asymptotic symmetries, for which the time $t$ is held fixed and the radial coordinate $r$ on $\Sigma$ is taken to infinity. This means that we do not consider the asymptotic symmetry group on the whole of spatial infinity of Minkowski spacetime,\footnote{Spatial infinity understood as the timelike boundary at which spacelike geodesics end connects the infinite past with the infinite future, and is therefore itself infinitely extended in time and not instantaneous.} but only at one instant. It is sometimes said that asymptotic analyses are more of an art than a science \cite[p. 34]{stromingerLecturesInfraredStructure2018}, but for the specific case of Yang-Mills theory on a Euclidean Cauchy surface we will present a reasonably algorithmic method for deriving the physical gauge group from the assumption of a well-defined instantaneous Lagrangian, which is a basic principle of classical field theory \cite{gotayMomentumMapsClassical2004a}. We expect that this method can be extended at least to Yang-Mills theory on Cauchy surfaces in other spacetimes than Minkowski, and likely also to the gravitational field itself.

The case of Maxwell theory, possibly with a Higgs field, on Euclidean space has been studied extensively in the foundations of physics community, see e.g. \cite{kossoEmpiricalStatusSymmetries2000,bradingAreGaugeSymmetry2004,healeyGaugingWhatReal2007,struyveGaugeInvariantAccounts2011a,tehGalileoGaugeUnderstanding2016,wallaceIsolatedSystemsTheireen,wallaceIsolatedSystemstwee,gomesHolismEmpiricalSignificance2021,gomesGaugingBoundaryFieldspace2019,gomesQuasilocalDegreesFreedom2021,berghoferGaugeSymmetriesSymmetry2023,gomesGaugeTheoryGeometrisation2025}. The terminology used there to describe physical and trivial gauge symmetries, respectively, is that of \textit{direct empirical significance} (DES) and \textit{redundant} gauge transformations \cite{tehGalileoGaugeUnderstanding2016}. Instead of `carrying DES' we will simply say `physical'. Redundant gauge transformations are contrasted with \textit{formal} gauge transformations, which are just the full infinite-dimensional gauge group $\mathcal{G}$ without any regard for their physical status. In the literature, the group of physical gauge symmetries for pure electromagnetism on $\Sigma$ with spatial asymptotic boundary conditions has been identified as the asymptotic symmetry group
\begin{align*}
    \mathcal{G}_\text{Phys}=\mathcal{G}^I/\mathcal{G}_0^\infty,
\end{align*}
where $\mathcal{G}^I$ denotes the subgroup of the formal gauge group $\mathcal{G}$ whose elements leave asymptotic boundary conditions  invariant\footnote{Hence the notation $I$, which will be used throughout to denote classes of maps that leave the asymptotic boundary conditions invariant, i.e. that are constant at infinity (except in the broken phase of the Yang-Mills-Higgs theory, where boundary-preserving transformations must actually vanish at infinity, see Section \ref{higgs}).} (the ``allowed'' symmetries), and $\mathcal{G}^\infty_0$ is the subgroup of redundant gauge transformations that are generated by the first-class constraints of the theory (the ``trivial'' symmetries). Here the $\infty$-superscript stands for the trivial action of these transformations at infinity\footnote{Throughout this article we use the \textit{subscript} $\infty$ to denote certain conditions at asymptotic infinity (usually the vanishing of classes of maps), which is not to be confused with the superscript denoting smoothness. Only for $\mathcal{G}^\infty_0$ have we used $\infty$ as a superscript since there we already have the subscript $0$ and there is no danger of confusion.} and the subscript $0$ denotes the identity component of $\mathcal{G}^\infty$. The identification of redundant gauge symmetries as the ones generated by the first-class\footnote{First-class constraints are constraints whose Poisson bracket with any other constraint is again a constraint, i.e. vanishes weakly. Not to be confused with primary constraints, which are obtained directly from the Legendre transform, without using the equations of motion.} constraints is based on the Dirac-Bergmann theory of constraints \cite{diracGeneralizedHamiltonianDynamics1950,PhysRev.83.1018}, in which one takes Poisson brackets of the first-class constraints with the fields of the theory to generate gauge transformations. For details see e.g. \cite{henneauxQuantizationGaugeSystems1992,lusannaDiracBergmannConstraintsPhysics2018,pittsFirstClassConstraint2014,pooleyFirstclassConstraintsGenerate2022,bradley2,bradleyRelationshipLagrangianHamiltonian2025}.

For electromagnetism on three-dimensional space $\Sigma$ the group $\mathcal{G}^I$ of boundary-preserving gauge transformations is identified as consisting of those transformations $g\colon \Sigma\to U(1)$ that become asymptotically constant \cite{tehGalileoGaugeUnderstanding2016,wallaceIsolatedSystemstwee}. Furthermore, the subgroup $\mathcal{G}_0^\infty$ is identified as the one generated by the Gauss law constraint, consisting of all transformations $g\colon \Sigma\to U(1)$ that asymptotically approach the identity \cite{balachandranGaugeSymmetriesTopology1994,tehGalileoGaugeUnderstanding2016}. The quotient is hen said to be isomorphic to $U(1)$ itself, i.e. the group of global (or \textit{rigid}) gauge symmetries \cite{Giulini:1994bi,struyveGaugeInvariantAccounts2011a,tehGalileoGaugeUnderstanding2016,wallaceIsolatedSystemstwee}. 

However, the derivations supporting these results are at the least shaky and sometimes simply incorrect. The common lore is that one must impose asymptotic fall-off boundary conditions on the spatial components of the gauge field, e.g.
\begin{align*}
    A_i\to 0+\mathcal{O}(r^{-2}),\;\;\;\;\; i=1,2,3,
\end{align*}
to ensure finiteness of energy and/or action \cite{lusannaDiracObservablesHiggs1997}. It is then said that the gauge group must preserve these conditions \cite{struyveGaugeInvariantAccounts2011a} and must therefore be restricted to $\mathcal{G}^I$. But this argument is problematic, since energy and action only depend on gauge-invariant quantities (the field strength tensor). Thus there is no need to require gauge fields to become zero asymptotically: we need only require that they become pure gauge, a point that was already noted by Atiyah \cite[Section I.4]{Atiyah:1979iu}. But any gauge transformation preserves this condition (of being pure gauge), so it would seem naively that one can always allow the full gauge group $\mathcal{G}$, instead of restricting to $\mathcal{G}^I$. There is an additional critique: the very statement $A_i\to 0$ is made in a specific gauge. What we call ``zero'' is therefore gauge-dependent. Thus, the fact that this asymptotic boundary condition is not preserved by most gauge transformations is not surprising - it is a consequence of our working in a gauge. If true, this would greatly enlarge the group $\mathcal{G}_\text{Phys}$, well beyond the group of global (rigid) gauge transformations.

One aim of this article is to explain why we must in fact still restrict to the subgroup $\mathcal{G}^I$, although this does not follow directly from finiteness of energy but from considering the structure of the domain of the instantaneous Lagrangian, which for a first-order theory on a spacetime without boundaries would be the tangent bundle to the instantaneous configuration space \cite{gotayMomentumMapsClassical2004a}. For finite-dimensional, non-covariant systems this latter fact is the foundation for proving the equivalence of the Euler-Lagrange equations and stationarity of action in variational principles, see e.g. \cite[Chapter 8]{marsdenIntroductionMechanicsSymmetry1999} or \cite[Chapter 19]{dasilvaLecturesSymplecticGeometry2008}. But gauge field theories are infinite-dimensional systems with infinite-dimensional symmetry groups, resulting in the added difficulty that the Lagrangian is degenerate (exhibits constraints) \cite{henneauxQuantizationGaugeSystems1992,binz}. In that case, not all vectors in the tangent bundle to configuration space admit solutions to the Euler-Lagrange equations of which they are the initial datum \cite[Section 6.4]{binz}. Nonetheless, the constraints are found primarily through the instantaneous Legendre transform $L\colon TQ\to T^*Q$ from the tangent bundle to the cotangent bundle of the instantaneous configuration space $Q$ \cite{gotayMomentumMapsClassical2004a}. The instantaneous constraint surface $\mathcal{C}$ is the image $L(TQ)$ of the tangent bundle under the instantaneous Legendre transform \cite{henneauxQuantizationGaugeSystems1992,binz,gotayMomentumMapsClassical2004a,Gotay2006MomentumMA}. Thus, in the instantaneous formulation of gauge field theories, one starts from an instantaneous Lagrangian defined on the tangent bundle to configuration space. Of course, one can instead work covariantly, but the theory must nonetheless have a well-defined 3+1 split with associated instantaneous Lagrangian \cite{gotayMomentumMapsClassical2004,gotayMomentumMapsClassical2004a}. This requirement translates boundary conditions on electric fields, required for finite energy, to the gauge fields themselves.

Besides the problem of correctly identifying $\mathcal{G}^I$, there is further obscurity in the literature when $\mathcal{G}_\text{Phys}$ is identified with the group of global (rigid) gauge symmetries. This pertains to the appropriate \textit{rate} at which transformations $g\in\mathcal{G}^I$ must become constant asymptotically, and the rate at which elements $g\in\mathcal{G}^\infty_0$ must approach the identity. It is only when these rates are exactly equal that we can conclude that the quotient of these two subgroups of $\mathcal{G}$ is isomorphic to $U(1)$ (in the Abelian case). However, in the usual approach it is not obvious that these rates are the same. To see this, note that, in 3-dimensional space the electric field must vanish asymptotically with order at least $\mathcal{O}(r^{-3/2-\epsilon})$ to guarantee that it is square-integrable,\footnote{Square-integrability is required because the energy carried by the electric field is the integral of the square of its norm, and this energy is required to be finite.} where $\epsilon>0$ is any (small) number. This same rate is then imposed on the gauge field itself, from which it is concluded that gauge transformations $g\colon \Sigma\to G$ must become constant asymptotically to preserve this boundary condition. But at what rate?  In the Abelian case, we would need the gauge parameter $\lambda\colon \Sigma\to\R$ to be such that its derivative $\partial_i\lambda$ falls off with order $\mathcal{O}(r^{-3/2-\epsilon})$, if it is to be boundary-preserving. But what does this imply for $\lambda$ itself? It is not obvious that we can simply conclude that $\lambda\to \text{const}+\mathcal{O}(r^{-1/2-\epsilon})$, i.e. that $\lambda$ falls off towards a constant with one power of $r$ fewer. Indeed, there are examples of functions which themselves vanish in a certain limit but whose derivative behaves very badly. Besides, as noted in \cite{struyveGaugeInvariantAccounts2011a,tehGalileoGaugeUnderstanding2016}, our choice of asymptotic behavior of the fields has a large effect on what ``allowed'' transformations are contained in $\mathcal{G}^I$, apparently making the derivation of the physical gauge group quite arbitary.

Similar issues arise when considering the order of asymptotic behavior for ``trivial'' transformations $g\in\mathcal{G}^\infty_0$. In fact, in the argument by Balachandran \cite{balachandranGaugeSymmetriesTopology1994}, which formed the basis for arguments in \cite{tehGalileoGaugeUnderstanding2016}, the requirement that $g\to 1$ asymptotically is based on the need for a certain boundary term to vanish in the calculation of a specific Poisson bracket. But this boundary term contains the electric field, and so its vanishing could also be guaranteed simply by requiring rapid enough asymptotic fall-off of the electric field, so that gauge transformations do not need to approach the identity to ensure that this Poisson bracket is well-defined. We will again run into this trade-off between fall-off on fields versus gauge transformations in Section \ref{mommapsection}, and resolve it in Section \ref{localizable}. At any rate, it is clear that quite a lot of fine-tuning of asymptotic behavior is needed to ensure that, in the end, the quotient $\mathcal{G}_\text{Phys}=\mathcal{G}^I/\mathcal{G}^\infty_0$ corresponds precisely to the group of global (rigid) gauge transformations. This fine-tuning is highly unsatisfactory.

These ambiguities contrast sharply with other characterizations of the special status of global gauge symmetries, from which it is obvious that it is precisely the global gauge group that stands apart from other gauge transformations. We mention three such characterizations.

First, in the formalization of gauge theories using fiber bundles connections live on a principal $G$-bundle $P\to \Sigma$.\footnote{Usually one would of course consider bundles over all of $M\cong\R\times\Sigma$, but we anticipate that we will be working in temporal gauge.} Gauge transformations correspond to bundle automorphisms $P\to P$. But in the Abelian case, there is clearly a special class of gauge transformations, namely the ones that are given by the global action $G\times P\to P$, which forms part of the very definition of a principal bundle. In the non-Abelian case the action $f\colon P\to P$ defined by $f(p)=ph_1$, for some $h_1\in G$, does not necessarily define a bundle automorphism. Equivariance might fail since $f(ph_2)=ph_2h_1$ is not necessarily the same as $f(p)h_2=ph_1h_2$ if $h_1,h_2$ do not commute. Yet the central elements of $G$ do define a bundle automorphism this way. A connection on $P$ is a choice of horizontal subspace at every point $p\in P$, and since the global action of $G$ (for $G$ Abelian) on $P$ is, by definition, perfectly vertical, it is not felt by the connections.

Second, in the symplectic formulation of gauge theories the global gauge group appears as the obstruction to the possibility of a smooth symplectic reduction \cite{SNIATYCKI1988291,sniatycskiboundarycondspatboundeddom,rielloEdgeModesEdge2021,rielloHamiltonianGaugeTheory2024}. To see this, recall that any Hamiltonian group action on a symplectic manifold can be used to define a momentum map (Definition \ref{mommap}) such that, if the group acts freely\footnote{The action of a group $H$ on a set $X$ is called free if  $h\cdot x=x$ for some $x\in X$ implies that $g$ is the identity.} and properly\footnote{The action of a topological group $H$ (such as a Lie group) acting by homeomorphisms on a topological space $X$ (such as a manifold) is called proper if the map $H\times X\to X\times X$ is proper. A map between topological spaces is called proper if the inverse image of a compact set is compact.} on the zero set of this momentum map, one can take a symplectic quotient \cite{Marsden:1974dsb,marsdenIntroductionMechanicsSymmetry1999,binz,dasilvaLecturesSymplecticGeometry2008,Diez:2024dts}. However, since global gauge transformations can be viewed as the constant maps $g\colon \Sigma\to G$, they do not act freely. Namely, in the Abelian case, a connection $A$ transforms as 
\begin{align*}
    A\to A+g^{-1}dg,
\end{align*}
so if $g$ is constant then $dg$ is zero, and \textit{any} connection will be a fixed point of the global gauge group action. This prevents the possibility of a smooth symplectic reduction.\footnote{To resolve this we could consider the group $\mathcal{G}_*$ of \textit{pointed} gauge transformations, i.e.~those transformations that are the identity at some arbitary fixed point $x_0\in \Sigma$. Then the only global transformation is the trivial one and the action of $\mathcal{G}_*$ is free, so that the symplectic reduction is a smooth space. This approach is pursued in \cite{belotSymmetryGaugeFreedom2003}. We could also consider so-called \textit{irreducible connections}, i.e.~connections for which the holonomy group acts irreducibly. The gauge group does act freely on the space of irreducible connections \cite{singerGeometryOrbitSpace1981}.} The symplectic quotient will instead be a stratified space \cite{Sjamaar1991Stratified}. In the non-Abelian case constant gauge transformations $g$ do act by conjugation $A\mapsto g^{-1}Ag$, (non-Abelian gauge bosons are charged and self-interact under the force they themselves transmit), but even then the central global gauge transformations still do not act freely.

Thirdly, but relatedly, Gomes and Riello have used horizontal symplectic geometry to identify the global gauge group as carrying a different empirical status from other gauge transformations \cite{gomesGaugingBoundaryFieldspace2019,gomesUnifiedGeometricFramework2019,gomesQuasilocalDegreesFreedom2021,gomesHolismEmpiricalSignificance2021,rielloEdgeModesEdge2021,rielloHamiltonianGaugeTheory2024}. In electromagnetism this is achieved by means of the Dirac dressing
\begin{align*}
    h[A](\textbf{x})=\int_\Sigma \frac{d^3y}{4\pi}\frac{\partial^i A_i(\textbf{y})}{|\textbf{x}-\textbf{y}|},
\end{align*}
which singles out the gauge-invariant component of the gauge field $A$ on 3-dimensional space $\Sigma$. This dressing corresponds to a projection onto the Coulomb gauge and is insensitive precisely to the global gauge transformations, as these do not change $A$. Clearly this is related to the previous point: the common idea is that (central) global gauge transformations do not change the gauge field, whereas these do change the global phase of matter fields.\footnote{For this reason they are used in so-called 't  Hooft beam splitter \cite{tHooft:1980aiu} constructions, see e.g. \cite{greavesEmpiricalConsequencesSymmetries2014}.}

Thus we arrive at the central goal of this article: unifying the various approaches to deriving precisely the global gauge group as the one carrying physical significance, by carefully considering the configuration space of Yang-Mills fields and their spatial asymptotic boundary conditions. Our approach is as follows. We first construct the instantaneous configuration space of gauge fields in Section \ref{configspace}, without working in a particular spatial gauge. In Section \ref{boundarycond} we then use this construction to define boundary conditions in Yang-Mills theory that are necessary to ensure the existence of the instantaneous Lagrangian, and we examine their consequence for the structure of theinstantaneous state and configuration spaces of gauge fields. Subsequently, we find the redundant gauge symmetries, i.e. those generated by the Gauss law constraint, in Section \ref{redundant}, finally giving us the quotient of physical transformations. Lastly, we study what happens when a Higgs field is added in Section \ref{higgs}, in which case we find different boundary conditions for the unbroken and broken phases.

\section{The configuration space of gauge fields}\label{configspace}

In this Section we study the configuration space of Yang-Mills theories. We do so without working in a particular \textit{spatial} trivialization, which is of paramount importance for conceptual clarity. After all, if we impose boundary conditions while already working in a specific trivialization, then it is not surprising that most gauge transformations violate this boundary condition. However, it is then unclear whether this violation is really problematic or just an artifact of our choice to work in a trivialization, and we should avoid this ambiguity. 

The results of this Section are a necessary prerequisite for understanding the main point of Section \ref{boundarycond}: that the need to restrict the allowed gauge transformations to $\mathcal{G}^I$, i.e. the subgroup of transformations that leave the boundary conditions invariant, comes not directly from the boundary conditions themselves, but rather from the fact that a vanishing electric field on the boundary makes the gauge field non-dynamical there, effectively leading to a Dirichlet boundary condition that must be respected by gauge transformations and gives the instantaneous state space the structure of a tangent bundle.

Throughout we assume a 3+1 split of flat spacetime into $\Sigma\times\R$, where $\Sigma\cong\R^3$, and work in the temporal gauge, thus setting $A_0=0$. Though this completely fixes the time-component of the gauge field, it does not restrict the spatial gauge freedom at all. This means that we do not consider gauge transformations in the temporal component of the gauge field, but only in its spatial components. We do this because we are ultimately interested in understanding the breaking of spatial gauge transformations in the Higgs mechanism. 

We consider a principal $G$-bundle $P\to \Sigma$, where the structure group $G$ is some compact matrix Lie group such as $U(1)$ or $SU(N)$, with Lie algebra $\text{Lie}(G)=\mathfrak{g}$. The structure group should not be confused with the gauge group $\mathcal{G}=\text{Aut}(P)$ of all gauge transformations. Note that, since we assume $\Sigma\cong\R^3$, every bundle on $\Sigma$ is automatically trivializable. Thus we know there exists a global section. Crucially, however, we do not \textit{actually} trivialize the bundle, as this would force us into a particular gauge, leading to the conceptual confusion referred to above. For complete clarity, we work with a trivializable but untrivialized bundle.

A gauge field in Yang-Mills theory is a connection on this bundle $P$, i.e. a choice of horizontal distribution in the tangent bundle $TP$. Equivalently, a gauge field can be viewed as a Lie algebra-valued 1-form on $P$, i.e. an element $A\in\Omega^1(P,\mathfrak{g})$, that is both $G$-equivariant and reproduces the Lie algebra generators of the fundamental vector fields\footnote{That is: $A(X_\xi)=\xi$ for all $\xi\in\mathfrak{g}$, where $X_\xi$ denotes the fundamental vector field in $\mathfrak{X}(P)$ generated by $\xi$ through the right action of $G$ on $P$.} \cite{hamiltonMathematicalGaugeTheory2017}. $G$-equivariance means that $r_h^*\circ A=\text{Ad}_{h^{-1}}\circ A$ for all $h\in G$, where $\text{Ad}:G\to\text{GL}(\mathfrak{g})$ denotes the adjoint representation\footnote{Defined by $\text{Ad}_h(X)=hXh^{-1}$, where $h\in G, X\in\mathfrak{g}$.} and $r_h^*\colon\mathfrak{g}\to\mathfrak{g}$ the pullback  of the right multiplication $r_h\colon G\to G$ by $h\in G$. Such a connection 1-form $A$ can be pulled down to $\Sigma$ if we choose a trivializing section $s\colon \Sigma\to P$, in which case it is acted upon by the gauge group $\mathcal{G}$ in the usual way:
\begin{align*}
    s^*\tilde{g}A=\tilde{g}^{-1}s^*A\tilde{g} +\tilde{g}^{-1}d\tilde{g},\;\;\;\;\; \tilde{g}\in C^\infty(\Sigma,G).
\end{align*}
Here $s^*\colon \Omega^1(P,\mathfrak{g})\to\Omega^1(\Sigma,\mathfrak{g})$ denotes the pullback through the section $s$, and we have used the isomorphism $\mathcal{G}=\text{Aut}(P)\cong C^\infty(\Sigma,G)$ induced by $s$, which sends $g\mapsto\tilde{g}$. The isomorphism between these two groups is as follows. If we have a $G$-valued map $g\colon \Sigma\to G$, then we can produce a bundle automorphism $f\colon P\to P$ using the section $s\colon \Sigma\to P$, namely by defining $$f(p)=p\cdot s(\pi(p)).$$ Henceforth we drop the tilde on $g$ for the sake of simplicity.

If we write $\text{Conn}(P)$ for the space of all connection 1-forms on $P$, then the space of ``coordinates'' and ``velocities'' in the instantaneous formulation of Yang-Mills theory naively equals the tangent bundle $T\text{Conn}(P)$ to $\text{Conn}(P)$ \cite{gotayMomentumMapsClassical2004a}. However, as we will see in Section \ref{boundarycond}, asymptotic boundary conditions are required on the tangent vectors (electric fields) in this tangent bundle, thereby complicating the construction. Now, such asymptotic boundary conditions are imposed as fall-off rates in reference to the space $\Sigma$, rather than the bundle $P$, so we need to bring down our fields to $\Sigma$ in order to define boundary conditions on them. We could do this by working in a trivialization, but we have just argued that it is vital to work with an untrivialized bundle. Fortunately, we can work directly on $\Sigma$ without the need to trivialize.

Recall that a $k$-form $\omega\in\Omega^k(P,\mathfrak{g})$ is called \textit{horizontal} if it vanishes whenever at least one vector it eats is vertical, i.e. if for all $p\in P$ we have $\omega_p(X_1,...,X_k)=0$ whenever $X_i\in V_pP=\text{ker}(\pi_*)$ for some $1\leq i\leq k$. Here $V_pP=\text{ker}(\pi_*)$ denotes the space of vertical vectors at the point $p$, which should be thought of as the vectors that lie along the fibers (which are isomorphic to $G$) of $P$. Furthermore, we say a $k$-form $\omega$ is $\textit{of type Ad}$ if $r_h^*\circ \omega=\text{Ad}_{h^{-1}}\circ\omega$ for any $h\in G$. We denote the set of horizontal $k$-forms of type Ad by $\Omega^k_\text{hor}(P,\mathfrak{g})^\text{Ad}$. It is then a well-known result \cite{hamiltonMathematicalGaugeTheory2017} that if $A,A'\in\Omega^1(P,\mathfrak{g})$ are two connection 1-forms, then $A-A'\in\Omega^1_\text{hor}(P,\mathfrak{g})^\text{Ad}$ and for any $\omega\in\Omega^1_\text{hor}(P,\mathfrak{g})^\text{Ad}$ we have that $A+\omega$ is a connection 1-form. For the curvature we have $F(A)\in\Omega^2_\text{hor}(P,\mathfrak{g})^\text{Ad}$. In other words: differences of connections as well as curvatures are horizontal forms of type Ad. This is extremely useful because of the following well-known theorem \cite{hamiltonMathematicalGaugeTheory2017}:

\begin{theorem}\label{curvatureadjointbundle}
    Let $\pi\colon P\to \Sigma$ be a principal $G$-bundle. Then $\Omega^k_\text{hor}(P,\mathfrak{g})^\text{Ad}$ and $\Omega^k(\Sigma,\text{Ad}(P))$ are canonically isomorphic as vector spaces through the pullback\footnote{Recall that for a fiber bundle $E\to N$ any map $f\colon M\to N$ induces a pullback bundle $f^*E\to M$. In this case the pullback (of the adjoint bundle) is the trivial vector bundle $P\times \mathfrak{g}$.} $\pi^*$.
\end{theorem}

Here $\text{Ad}(P)$ denotes the adjoint bundle.\footnote{The adjoint bundle is the associated real vector bundle $\text{Ad}(P)=P\times_\text{Ad}\mathfrak{g}$ constructed through the adjoint representation $\text{Ad}\colon G\to\text{GL}(\mathfrak{g})$. Here the product $P\times_\rho\mathfrak{g}$ signifies that we quotient $P\times\mathfrak{g}$ by the equivalence relation $(p,X)\sim (ph,\text{Ad}_{h^{-1}}(X))$ for $h\in G$.} Thus, if we choose a basis connection $A_\text{ref}$, we can view the space of connection 1-forms $\text{Conn}(P)$ as the vector space $\Omega^1(\Sigma,\text{Ad}(P))$. In other words: we can view differences of connections as well as curvatures as $\mathfrak{g}$-valued forms on $\Sigma$ instead of on $P$ \textit{without trivializing the bundle}, as long as we remember that it is in reference to the basis connection $A_\text{ref}$. For an Abelian structure group the adjoint bundle $\text{Ad}(P)$ is even trivial, i.e. just $\text{Ad}(P)=\Sigma\times\mathfrak{g}$, so that the space of connections becomes simply $\Omega^1(\Sigma,\mathfrak{g})$. 

Now, we know what the tangent space to a vector space looks like: it is isomorphic to the original vector space, even in infinite dimensions. This allows us to obtain the tangent bundle to the space of connections. We find $$T\text{Conn}(P)\cong T\Omega^1(\Sigma,\text{Ad}(P))\cong \Omega^1(\Sigma,\text{Ad}(P))\times \Omega^1(\Sigma,\text{Ad}(P)).$$ In electromagnetism $\mathfrak{g}=i\R$, so that $T\text{Conn}(P)$ simply equals $\Omega^1(\Sigma)\times\Omega^1(\Sigma)$. Before we end this Section, it is crucial to stress the following point.

\begin{remark}\label{affineremark}
    Two connections $A,A'\in\Omega^1(P,\mathfrak{g})$ cannot be added or subtracted. Of course, we can define $A+A'$ and $A-A'$ \textit{as forms}, but these forms do not reproduce the generators $\xi\in\mathfrak{g}$ of fundamental vector fields $X_\xi$, since $(A+A')(X_\xi)=2\xi$ and $(A-A')(X_\xi)=0$. Thus connection 1-forms cannot be added or subtracted to produce new connection 1-forms. By Theorem \ref{curvatureadjointbundle}, however, \textit{differences} of connections w.r.t. a fixed reference connection $A_\text{ref}$ form a vector space. Therefore differences of connections w.r.t. $A_\text{ref}$ can indeed be added and subtracted. When one trivializes $P$ with a section $s$, there is a preferred reference connection, namely the zero connection on the trivial bundle. On an untrivialized bundle, however, there is no notion of ``zero'', implying that the space of connections is an affine space. Thus, $T\text{Conn}(P)\cong T\Omega^1(\Sigma,\text{Ad}(P))$ is really an identification of the tangent bundles of affine spaces, where the origin of $\Omega^1(\Sigma,\text{Ad}(P)$ is ``forgotten''.
\end{remark}

\section{Asymptotic boundary conditions and the gauge group}\label{boundarycond}

Thus far we have not considered any boundary conditions on the connection 1-forms or the tangent vectors in $T\text{Conn}(P)\cong T\Omega^1(\Sigma,\text{Ad}(P))$, nor on the curvatures of the connections, even though this is essential for ensuring existence of the instantaneous Lagrangian. The latter is an integral of the instantaneous Lagrangian density \cite{gotayMomentumMapsClassical2004a} over $\Sigma\cong\R^3$, so terms that appear in the density must fall off asymptotically with order at least $\mathcal{O}(r^{-3-\epsilon})$ to make this integral well-defined. To understand precisely what these terms are we first derive the instantaneous Lagrangian of Yang-Mills theory in temporal gauge from the covariant Lagrangian. 

Our goal in this Section is then to identify a subspace $Q\subset \Omega^1(\Sigma,\text{Ad}(P))$ of the space of all gauge fields, that is such that the curvatures of its elements as well as the vectors in its tangent bundle $TQ$ satisfy the asymptotic boundary conditions required to make the instantaneous Lagrangian well-defined.

\subsection{Boundary conditions on Yang-Mills fields}\label{boundaryconditionssection}

In GiMmsy's\footnote{GiMmsy stands for Gotay, Isenberg, Marsden, Montgomery, Sniatycki en Yasskin, with the names of the ``main protagonists'' capitalized \cite{blohmann}.} instantaneous formulation of field theories \cite{binz,gotayMomentumMapsClassical2004,gotayMomentumMapsClassical2004a,Gotay2006MomentumMA}, one moves from the covariant to the canonical theory by implementing a 3+1 split of spacetime and defining the associated \textit{instantaneous Lagrangian} and \textit{instantaneous Legendre transform}. For any first-order theory, the instantaneous Lagrangian is a map $\mathcal{L}\colon TQ\to\R$ \cite{gotayMomentumMapsClassical2004a}. Elements of $TQ$ consist of pairs $(A,\alpha_A)\in Q\times T_AQ$ of gauge fields and tangent vectors. We think of $\alpha_A$ as the electric field, but it is entirely independent of $A$ as long as we do not impose the equations of motion, which is one reason why we have chosen not to use the symbol $E$ (the other reason is that we will use $E$ to denote the conjugate momentum to $A$ in Section \ref{redundant}) . The tangent vectors $\alpha_A$ are the ``velocities'' at the ``coordinate'' $A$. 
\begin{proposition}
The instantaneous Lagrangian of Yang-Mills theory in temporal gauge is \cite{armsLinearizationStabilityGravitational1979}
\begin{align}\label{YMlagrangian}
    \mathcal{L}(A,\alpha_A)=\frac{1}{2}\norm{\alpha_A}^2-\frac{1}{2}\norm{F(A)}^2.
\end{align}
Here $F(A)$ denotes the curvature 2-form of the connection 1-form $A$, which is the magnetic field, and $\norm{\cdot}$ is the usual norm on forms:
\begin{align*}
    \norm{\omega}^2=\int_\Sigma \text{Tr}(\omega \wedge *\omega),
\end{align*}
where $*$ denotes the Hodge star operator.\footnote{It is important to remember for our conformal analysis in Section \ref{conformal} that the Hodge star operator ``contains'' the metric.} 
\end{proposition} 
\textit{Proof.} We can derive expression \eqref{YMlagrangian} for the Lagrangian from the usual covariant action on spacetime $M=\Sigma\times\R$:
\begin{align*}
    \mathcal{S}(\tilde{A})=-\frac{1}{2}\int_M\text{Tr } F(\tilde{A})\wedge * F(\tilde{A})=-\frac{1}{2}\int_\R \int_\Sigma\text{Tr }F(\tilde{A})\wedge * F(\tilde{A}).
\end{align*}
Here we have written $\tilde{A}$ to stress that this is a gauge field on spacetime $M$ instead of space $\Sigma$. Denoting coordinates on $\Sigma$ by $x^i$ and the coordinate on $\R$ by $t=x^0$, it is not difficult to show that the action in coordinates becomes the usual \cite{hamiltonMathematicalGaugeTheory2017}
\begin{align*}
    \mathcal{S}(\tilde{A})=-\frac{1}{4}\int_\R dt\int_\Sigma dx^3\text{ Tr}(F_{\mu\nu}F^{\mu\nu})=-\frac{1}{4}\int_\R dt\int_\Sigma dx^3\text{ Tr}\left( 2F_{0i}F^{0i} +F_{ij}F^{ij}\right),
\end{align*}
where $\mu=0,1,2,3$, $i=1,2,3$ and $F_{\mu\nu}=\partial_\mu A_\nu-\partial_\nu A_\mu+[A_\mu,A_\nu]$ is the antisymmetric field strength tensor (which clearly satisfies $F_{00}=0$). The term $\text{Tr}(F_{0i}F^{0i})$ is (minus) the energy of the electric field (the ``kinetic'' energy), and the term $\text{Tr}(F_{ij}F^{ij})$ equals twice the energy of the magnetic field (the ``potential'' energy). 

If we now impose temporal gauge $A_0=0$ we obtain $F_{0i}=\partial_0 A_i=\dot{A}_i$. We can then rewrite the action as\footnote{We use $(-,+,+,+)$ signature for the metric, which explains the minus sign in $-\dot{A}_i\dot{A}^i$.}
\begin{align*}
    \mathcal{S}(\tilde{A})=\frac{1}{4}\int_\R dt\int_\Sigma dx^3\text{ Tr}\left(-2\dot{A}_i \dot{A}^i-F_{ij}F^{ij}\right)=\int_\R dt\; \mathcal{L}(A_i,\dot{A}_i).
\end{align*}
But $F_{ij}$ is just the curvature of the connection $A_i$ on three-dimensional space $\Sigma$, so in coordinate-free notation we find, with slight abuse of notation:
\begin{align*}
    \mathcal{S}(\tilde{A})=\frac{1}{2}\int_\R dt\int_\Sigma\text{Tr}\left(\dot{A}\wedge *\dot{A}-F(A)\wedge * F(A)\right)=\frac{1}{2}\int_\R dt\left(\norm{\dot{A}}^2-\norm{F(A)}^2\right),
\end{align*}
where it is understood that $A\in Q\subset \Omega^1(\Sigma,\text{Ad}(P))$ signifies the spatial part of $A_\mu$. Technically there is no sense in which the spatial gauge field $A$ has a time-derivative $\dot{A}$. What is really meant by this expression is that $\dot{A}$ should be viewed as a tangent vector at the point $A$, obtained as a derivative along a curve $\R\to Q$. Thus we replace $\dot{A}\to \alpha_A\in T_AQ$ and we obtain the Lagrangian in Eq. \eqref{YMlagrangian}. $\Box$
\\
\\
\indent Now, we need the instantaneous Lagrangian to be well defined as an integral over $\Sigma$, and so we require both $\norm{\alpha_A}$ and $\norm{F(A)}$ to be separately finite, anticipating also that the energy is the sum of these. As these norms are just integrals over 3-dimensional space, square-integrability requires that $\alpha_A$ and $F(A)$ fall-off sufficiently quickly towards spatial asymptotic infinity. Denoting by $g_\Sigma$ the metric and writing
\begin{align*}
    \omega\wedge *\omega= g_\Sigma(\omega,\omega)\,d\text{Vol}_{g_\Sigma},\;\;\;\;\;\omega\in\Omega^k(\Sigma,\text{Ad}(P)),
\end{align*}
we need to require that, as $r\to\infty$:
\\

(i) $g_\Sigma(\alpha_A,\alpha_A)\to 0+\mathcal{O}\left(r^{-3-\epsilon}\right)$;

(ii) $g_\Sigma\left(F(A),F(A)\right)\to 0+\mathcal{O}\left(r^{-3-\epsilon}\right)$,
\\
\\
where $\epsilon>0$ is a small number. Since in coordinates we have (using Einstein summation convention)
\begin{align*}
    &g_\Sigma(\alpha_A,\alpha_A)=g_\Sigma^{ij}(\alpha_A)_i(\alpha_A)_j;
    \\
    &g_\Sigma(F(A),F(A))=g_\Sigma^{ik}g_\Sigma^{jl}F(A)_{kl}F(A)_{ij},
\end{align*}
we find the following boundary conditions in Cartesian coordinates:
\\

(i) $(\alpha_A)_i\to 0+\mathcal{O}\left(r^{-3/2-\epsilon}\right)$;

(ii) $F(A)_{ij}\to 0+\mathcal{O}\left(r^{-3/2-\epsilon}\right)$.
\\
\\
In spherical coordinates, however, the inverse of the metric $g_\Sigma=dr^2+r^2d\Omega^2$ gives a factor $r^{-2}$ for each of the angular coordinates (and a factor $r^{-4}$ for the double angular coordinate $F_{\theta\phi}$), leading to the boundary conditions:
\\

(i) $(\alpha_A)_r\to 0+\mathcal{O}\left(r^{-3/2-\epsilon}\right),\;\;\;\;(\alpha_A)_\theta,\;\;(\alpha_A)_\phi\to 0+\mathcal{O}\left(r^{-1/2-\epsilon}\right)$;

(ii) $F(A)_{r\theta},\;\;F(A)_{r\phi}\to 0+\mathcal{O}\left(r^{-1/2-\epsilon}\right),\;\;\;\;F(A)_{\theta\phi}=\mathcal{O}\left(r^{1/2-\epsilon}\right)$.
\\
\\
\indent All in all we see that the gauge field $A$ must become flat at asymptotic infinity sufficiently quickly and the tangent vector ``electric field'' $\alpha_A$ must vanish at infinity. We note that there is apparently no requirement for the gauge field itself to vanish at infinity, since it does not appear in the Lagrangian directly. It only needs to become flat \cite{Atiyah:1979iu}. But this raises the question: do the above boundary conditions produce an appropriate domain for the instantaneous Lagrangian, i.e. a tangent bundle? That is: if we take $Q$ to consist of those connections that become flat asymptotically at the rate indicated above, will its tangent space $T_AQ$ at a point $A\in Q$ then consist precisely of those $\alpha_A$ that approach zero asymptotically at that same rate? The answer is no. To see this, consider the space of flat connections at infinity and examine its tangent space. It should consist of the zero vector only, since we require $\alpha_A$ to vanish at infinity. In other words: there are no dynamical degrees of freedom at infinity, since the velocities vanish there. Thus, the gauge field is frozen at infinity: it can never change its value there. This leads to something akin to superselection sectors, labelled by all possible gauge field configurations at infinity. Each sector has the structure of a tangent bundle, but one cannot move between the sectors, because the gauge field is non-dynamical on the asymptotic boundary.

There are two obvious ways of handling this ``superselection structure''. Either we account for all sectors together, while keeping track of the fact that one cannot dynamically move between them, or we choose to work in one sector. In the first case the instantaneous state space has the structure of a disjoint union of tangent bundles, each presenting a sector defined by a gauge field configuration on the boundary. In this article, however, we choose the latter option, restricting ourselves to one dynamical sector, which equals the tangent bundle to the instantaneous configuration space of gauge fields approaching a fixed configuration at infinity, i.e. satisfying a Dirichlet boundary condition. In a future work \cite{borsboominstantaneous} we will pursue the former option and study the instantaneous state space with asymptotic boundary conditions as a ``stratified'' space, with the ``strata'' labeled by the possible values of the gauge field on the boundary.

In order to restrict the system to one dynamical sector, we need to restrict $Q$ so that it consist of just a single configuration at infinity, i.e. some \textit{fixed} asymptotic boundary choice of flat connection at infinity. This means that we impose a flat asymptotic Dirichlet boundary condition at infinity. Gauge transformations must leave this fixed choice of flat connection at infinity invariant.

\subsection{Conformal analysis}\label{conformal}

Having motivated the necessity of choosing a specific asymptotic Dirichlet boundary condition (rather than working simultaneously with all dynamical sectors), we must be more precise about the rates at which the boundary condition is approached.The conformal invariance of Yang-Mills theory allows one to make use of a conformal embedding of Minkowski spacetime $(M,\eta)$ into a Lorentzian manifold $(\hat{M},\hat{\eta})$ with compact Cauchy surfaces. Such an embedding is a map $f\colon (M,\eta)\to(\hat{M},\hat{\eta})$ which sends $M$ to the interior of $\hat{M}$ and which is such that $f^*\hat{\eta}=K^2\eta$ for some positive function $K$, the conformal factor. We can take $\hat{M}$ to be $\R\times S^3$ with the metric $\hat{\eta}=-d\tau^2+g_{S^3}$. Using standard angular coordinates $(\alpha,\beta,\gamma)$ for $S^3$ and spherical coordinates $(r,\theta,\phi)$ for $\R^3$, we have
\begin{align*}
    g_{S^3}=d\alpha^2+\sin^2(\alpha)\left(d\beta^2+\sin^2(\beta) d\gamma^2\right),
\end{align*}
and the embedding $f\colon\R\times\R^3\to\R\times S^3$ is explicitly given by $\tau\circ f=\arctan(t+r)+\arctan(t-r)$, $\alpha\circ f=\arctan(t+r)-\arctan(t-r)$, $\beta\circ f=\theta$ and $\gamma\circ f=\phi$ \cite[p. 384]{binz}. This gives the conformal factor
\begin{align*}
    K^2=\frac{4}{((t+r)^2+1)((t-r)^2+1)},
\end{align*}
which at fixed $t$ clearly satisfies $K\to 0+\mathcal{O}(r^{-2})$. 

One can attach the sphere of directions at spatial infinity \cite{SNIATYCKI1988291} such that $\hat{M}$ has the structure of a manifold with boundary $\partial\hat{M}$ on which $K$ vanishes \cite[Proposition 8.5.2]{binz}. The Cauchy surface $\Sigma\cong \R^3$ is then mapped into the interior of a compact space $\hat{\Sigma}$ with boundary $\partial\hat{\Sigma}\cong S^2$ (the celestial sphere of directions at infinity), so we can view asymptotic infinity of $\Sigma$ as $S^2$. For simplicity we consider $\Sigma$ at time $t=0$, so that the conformal embedding $\Sigma\to\hat{\Sigma}$ reduces to $R=2\,\text{arctan}(r),\theta=\theta,\phi=\phi$, where $0\leq R\leq \pi$ denotes the radial coordinate on $\hat{\Sigma}$ and $\theta,\phi$ the angular coordinates on both $\Sigma$ and $\hat{\Sigma}$. It is readily verified\footnote{Using the relation $\text{sin}^2(2\,\text{arctan}(r))=\left(\frac{2r}{1+r^2}\right)^2=\frac{4r^2}{(1+r^2)^2}$.} that the metric $g_{\hat{\Sigma}}=dR^2+\text{sin}^2(R)d\Omega^2$ is pulled back to $4(1+r^2)^{-2}\left(dr^2+r^2 d\Omega^2\right)$, i.e. the Euclidean metric on $\Sigma$ with conformal factor $K=2(1+r^2)^{-1}$. The asymptotic boundary then corresponds to $R=\pi$.

Since gauge fields are objects that are defined without reference to the metric, they do not transform with any conformal factor when considered on $\hat{\Sigma}$ \cite{Sniatycki:1993wb}. Electric fields do, if they are defined using the metric, transform as $E\to K^{-1}E$ (guaranteeing that there can be a nonzero electric flux at infinity), though this point is a bit subtle and we will come back to it in Section \ref{electricfluxsection}. However, component functions of forms may transform with conformal factors. This is most easily seen in spherical coordinates. Indeed, let $\alpha\in\Omega^1(\Sigma,\text{Ad}(P))$ and write $\alpha=\alpha_r dr+\alpha_\theta d\theta+\alpha_\phi d\phi$. Suppose that $\alpha=f^*\hat{\alpha}$ for some $\hat{\alpha}\in\Omega^1(\hat{\Sigma},\text{Ad}(\hat{P}))$. Then what asymptotic behavior on $\alpha$ is implied by the fact that $\hat{\alpha}$ extends smoothly to the conformal boundary $\partial\hat{\Sigma}$? Writing $\hat{\alpha}=\hat{\alpha}_R dr+\hat{\alpha}_\theta d\theta+\hat{\alpha}_\phi d\phi$, it is clear that we will simply have $\hat{\alpha}_\theta=\alpha_\theta$ and $\hat{\alpha}_\phi=\hat{\alpha}_\phi$, since the conformal embedding $f|_\Sigma\colon \Sigma\to\hat{\Sigma}$ does not change the angular coordinates. For the radial coordinate, however, we have $$dR=d(2\text{arctan}(r))=2dr/(1+r^2),$$ so that $f^*\hat{\alpha}_R=K\,\alpha_r$. This means that, through the conformal embedding, the radial coordinate function of a 1-form is asymptotically suppressed by $r^{-2}$.

Since the angular coordinates are left invariant, however, the assumption that $\hat{\alpha}$ smoothly extends to the boundary $\partial\hat{\Sigma}$ implies for the pulled back angular components that $\alpha_\theta,\alpha_\phi=\mathcal{O}(1)$. But this is not enough for square integrability on $\Sigma$ since, as we saw before, that requires at least $\alpha_\theta,\alpha_\phi=\mathcal{O}(r^{-1/2-\epsilon})$. Thus, we need to impose the additional requirement that $\hat{\alpha}|_{\partial\hat{\Sigma}}=0$.
\begin{proposition}
    The condition $\hat{\alpha}|_{\partial\hat{\Sigma}}=0$ guarantees square-integrability.
\end{proposition}

\noindent\textit{Proof.} Since $\hat{\alpha}$ is smooth even on the boundary $\partial\hat{\Sigma}$, we can perform a Taylor expansion at a point on the boundary in the coordinate $\rho=\pi-R\geq 0$ around $\rho=0$. Expanding
\begin{align*}
    &\hat{\alpha}_\theta=a_0+a_1\rho+a_2\rho^2+...,
    \\
    &\hat{\alpha}_\phi=b_0+b_1\rho+b_2\rho^2+...,
\end{align*}
we see that the requirement $\hat{\alpha}|_{\partial\hat{\Sigma}}=0$ implies $a_0=b_0=0$, so that to lowest order $\hat{\alpha}_\theta$ and $\hat{\alpha}_\phi$ are linear in $\rho$. But for large $r$:
\begin{align*}
    \rho=\pi-R=\pi-2\,\text{arctan}(r)=\pi-2\left(\frac{\pi}{2}-\frac{1}{r}+\frac{1}{3r^3}+\mathcal{O}(r^{-5})\right)=\frac{2}{r}+\mathcal{O}(r^{-3}).
\end{align*}
Since $\hat{\alpha}_\theta,\hat{\alpha}_\phi\sim \rho$ close to the boundary $\partial\hat{\Sigma}$, this implies that $f^*\hat{\alpha}_\theta,f^*\hat{\alpha}_\phi\to 0+\mathcal{O}(r^{-1})$. So the requirement that a 1-form $\alpha$ on $\Sigma$ equals the pullback of some 1-form $\hat{\alpha}$ on $\hat{\Sigma}$ that vanishes on the boundary, gives the conditions $\alpha_r\to 0+\mathcal{O}(r^{-2})$ and $\alpha_\theta,\alpha_\phi\to 0+\mathcal{O}(r^{-1})$, which are strong enough to guarantee square-integrability. $\Box$
\\
\\
\indent In a similar fashion we can analyze what asymptotic behavior is implied for a 2-form $F\in\Omega^2(\Sigma,\text{Ad}(P))$ if we assume it to come from some $\hat{F}\in\Omega^2(\hat{\Sigma},\text{Ad}(\hat{P}))$ that smoothly extends to the boundary $\partial\hat{\Sigma}$. 
\begin{proposition}
    A curvature 2-form on $\Sigma$ obtained as $f^*\hat{F}$, where $\hat{F}$ extends smoothly to $\partial\hat{\Sigma}$, is square-integrable.
\end{proposition}

\noindent\textit{Proof.} Since $F_{ii}$ components vanish by antisymmetricity, it suffices to check $F_{r\theta},F_{r\phi}$ and $F_{\theta\phi}$. As the radial component is the only one whose exterior derivative transforms with a conformal factor (i.e. $dR=K\,dr$), we find that $f^*\hat{F}_{r\theta}=K\,F_{r\theta}$, and the same for $F_{r\phi}$. The component $F_{\theta\phi}$ is left invariant. Thus, if a 2-form $F$ is assumed to equal a pullback $f^*\hat{F}$, then we automatically find that $F_{r\theta},F_{r\phi}\to 0+\mathcal{O}(r^{-2})$ and $F_{\theta\phi}=\mathcal{O}(1)$. Comparing to Section \ref{boundaryconditionssection} we find that this is easily enough to guarantee that $F$ is square-integrable. Thus, if one assumes a 1-form $A$ on $\Sigma$ to equal the pullback of a 1-form $\hat{A}$ on $\hat{\Sigma}$, then the curvature $F(\hat{A})$ is a well-defined 2-form on $\hat{\Sigma}$ and therefore one automatically finds that  $F(A)$ is square-integrable. $\Box$

\subsection{Boundary-preserving gauge transformations}

Applying the conformal analysis above to a Yang-Mills field $A\in Q$ with velocity tangent vector $\alpha_A\in T_AQ$ and curvature $F(A)$, we find that we need to assume that $\hat{\alpha}_{\hat{A}}\in T_{\hat{A}}\hat{Q}$ vanishes on the conformal boundary $\partial\hat{\Sigma}$, whereas we do not need to assume any behavior on $F(\hat{A})$ other than smoothness on $\hat{\Sigma}$. But since tangent vectors in $T_{\hat{A}}\hat{Q}$ are required to vanish on $\partial\hat{\Sigma}$, we are in the situation outlined in Section \ref{boundaryconditionssection}: the degrees of freedom on the boundary are non-dynamical, leading to disjoint sectors defined by the configuration of the gauge field on the boundary. As indicated earlier, we choose to work in one such sector by choosing a particular Dirichlet boundary condition, denoted $\hat{A}_\infty$, on the asymptotic boundary $\partial\hat{\Sigma}$, rather than keeping track of all sectors simultaneously. This ensures that the tangent space to the configuration space on the boundary is 0-dimensional, as required, so that the full instantaneous state space is simply given by the tangent bundle $T\hat{Q}$, where $\hat{Q}$ consists of all those gauge fields that agree with our choice of boundary condition.

However, such a choice of a fixed connection at infinity obviously \textit{breaks gauge invariance}, but in a trivial sense: any gauge transformation that is not constant\footnote{One may worry that the identification of $\text{Aut}(\hat{P}|_{\partial\hat{\Sigma}})$ with maps $\partial\hat{\Sigma}\to G$ fails because trivializability of $P\to \Sigma$ does not imply trivializability over the conformal boundary. However, the extension of $\hat{P}\to f(\Sigma)=\text{int}(\hat{\Sigma})$ to the boundary $\partial\hat{\Sigma}$ is a matter of choice, and we simply choose it to be trivializable.} at infinity will change $\hat{A}_\infty$. In the Abelian case, the group of gauge transformations that do preserve $\hat{A}_\infty$ consists precisely of all transformations that are constant at infinity. In the non-Abelian case one has to take into account the fact that even constant transformations may change $\hat{A}_\infty$ by means of a conjugation. The orbit of a connection under the conjugation action of the group of constant gauge transformations is itself a smooth manifold (whose dimension depends on $G$), with tangent vectors which are nonzero unless the boundary connection is invariant under $\text{Ad}(G)$. Intuitively, this corresponds to the fact that non-Abelian gauge fields carry currents even in the absence of matter fields. To avoid such currents at infinity we pick a boundary connection which is invariant under $\text{Ad}(G)$, e.g. zero. Then the asymptotically constant gauge group will leave this boundary choice invariant. This choice of picking the zero connection at infinity so that the full asymptotically constant gauge group is allowed, rather than allowing for any flat connection but only the central asymptotically constant transformations, harmonizes with Doplicher-Haag-Roberts superselection theory in algebraic QFT, in which the global gauge group $G$ gives rise to observable superselection sectors and can in turn be reconstructed from such a superselection structure \cite{doplicherFieldsObservablesGauge1969,doplicherFieldsObservablesGauge1969a,doplicherLocalObservablesParticle1971,doplicherNewDualityTheory1989,doplicherWhyThereField1990,haagAlgebraicApproachQuantum1964}.

In this way, we again arrive at the familiar fact that the group of boundary-preserving ``allowed'' gauge transformations $\mathcal{G}^I$ consists of those that become constant at infinity at the appropriate rate. We need not worry anymore about what this rate is precisely,\footnote{But note that, by choosing to formalize asymptotic infinite in the way done here, we picked $\mathcal{O}(r^{-2})$ as the preferred fall-off rate.} since it does not play a role when working on the compact space $\hat{\Sigma}$, where there is only one simple condition on the transformations in $\mathcal{G}^I$, namely that they are constant on $\partial\hat{\Sigma}$. It is also clear why gauge fields $A$, when viewed on $\hat{\Sigma}$, automatically approach the fixed flat connection $\hat{A}_\infty\in\Omega^1(\partial\hat{\Sigma},\mathfrak{g})$ at the same rate that tangent vectors $\alpha_A$ approach zero. This follows from understanding the space of ``electric fields'' as the tangent space to $Q$, a consequence of our choice to work in one specific dynamical boundary sector. Since $T_AQ\cong Q$ (viewed as vector spaces rather than affine spaces), any choice of asymptotic behavior for elements in $T_AQ$ automatically translates this behavior onto $Q$ itself.

\begin{remark}
    When choosing the \textit{zero connection} as the Dirichlet boundary condition on the conformal boundary, we are already viewing the space $\Omega^1(\partial\hat{\Sigma},\mathfrak{g})$ of connections on the boundary as a \textit{vector} space rather than an \textit{affine} space. Indeed, as an affine space, this space of connections has no zero element. It is therefore more precise to say that one first chooses a Dirichlet boundary condition as an element $\hat{A}_\infty\in \text{Conn}(\hat{P}_{\partial\hat{
    \Sigma}})$, i.e. as a connection 1-form over the boundary, and then one uses this connection as the reference appearing in Theorem \ref{curvatureadjointbundle}. This ensures that, on the boundary, the Dirichlet boundary condition appears as the zero connection. On the interior of space, the space of connections must then still be viewed as an affine space (cf. Remark \ref{affineremark}), but this procedure nonetheless gives a trivialization-independent definition of the dynamical sector defined by one's choice of Dirichlet boundary condition.
\end{remark}

\section{Redundant gauge symmetries and constraints}\label{redundant}

Having reproduced the result that, as long as one picks an $\text{Ad}(G)$-invariant boundary gauge field configuration, the subgroup of boundary-preserving gauge transformations $\mathcal{G}^I$ consists of those transformations that become constant at infinity - interpreted properly as the boundary of the compact space $\hat{\Sigma}$ - it is time we turn to the question of the redundancy or ``triviality'' of these gauge transformations. That is: which elements of $\mathcal{G}^I$ are generated by the Gauss law constraint, which is the first-class constraint defining the constraint surface of Yang-Mills theory,\footnote{Besides the $\Pi^0=0$ constraint that tells us that the time-component $A_0$ of the gauge field is a Lagrange multiplier, but which is excluded in our analysis because we are working in temporal gauge $A_0=0$ from the beginning.} and can therefore be interpreted to be unphysical?

In constrained Hamiltonian analysis, \textit{gauge orbits} are null directions\footnote{A symplectic form is required to be non-degenerate only on the full phase space and not on the constraint surface.} of the symplectic form pulled back to the constraint surface $\mathcal{C}$ \cite{henneauxQuantizationGaugeSystems1992}. These null directions give a clear definition of ``gauge'' in the redundant sense: they are not felt by the symplectic form, which is the central object in the classical structure of the theory. It was Dirac's great insight that these gauge orbits are generated by the first-class constraints \cite{diracGeneralizedHamiltonianDynamics1950,Dirac1964Lectures}. In symplectic geometry, this idea is made precise by means of the momentum map, which formalizes infinitesimally generated symmetries. Indeed, in Yang-Mills theory the constraint surface equals the inverse image of zero under the momentum map for the group of redundant, trivial gauge symmetries \cite{binz}. Thus, in order to discover precisely which transformations in $\mathcal{G}^I$ are redundant (trivial), we pursue the following strategy: we calculate for which infinitesimal gauge transformations the momentum map is given by the Gauss law constraint. This approach can be seen as a precise version of the argument from \cite{balachandranGaugeSymmetriesTopology1994} and is also pursued for finite boundaries in \cite{rielloHamiltonianGaugeTheory2024}, and we will follow it now to highlight how local and global gauge symmetries obtain a different physical status: only the former have the Gauss law constraint as their momentum map and should therefore be viewed as redundant.

However, we will then explain the weakness of such an approach in the setting of asymptotic boundaries: we run into the same issues about the appropriate rates of asymptotic behavior as those highlighted in the introduction. Thus we will be forced to revert back to the compact space $\hat{\Sigma}$ related to $\Sigma$ through a conformal embedding. We will then see that the redundant gauge transformations are the ones that equal the identity on the conformal boundary $\partial\hat{\Sigma}$, though understanding the behavior of the electric field on $\partial\hat{\Sigma}$ is a bit tricky. Subsequently we generalize this result by explaining the notion of an \textit{infinitesimal localizable symmetry}, which in the mathematical literature are the symmetries that yield Noether's second theorem and the resulting constraints, and are therefore redundant. Only global gauge symmetries are not localizable, so these should be viewed as carrying a different empirical status than local gauge symmetries. They are symmetries that do not lead to constraints, similar to e.g. rotational symmetry for a point particle moving in Euclidean space.\footnote{The Lagrangian for such a particle is $\mathcal{L}(q,v)=\frac{1}{2}g_q(v,v)-V(q)$, where $g_q\colon T_q \R^3\times T_q\R^3\to\R$ is a metric. The symmetries of the system are the isometries that leave the potential $V$ invariant. If the potential is rotationally symmetric, then rotations are symmetries. But the Legendre transform $L\colon T\R^3\to T^*\R^3$ is given by $v_q\to g_q(v,\cdot)$, which is clearly a diffeomorphism. This means that there are no constraints.}

\subsection{The momentum map for the gauge group}\label{mommapsection}

Let $Q_{A_\infty}\subset\Omega^1(\Sigma,\text{Ad}(P))$ denote the space of connections on $P\to \Sigma$ satisfying the boundary conditions we arrived at in the previous Section, i.e. approaching a fixed flat\footnote{Though, as we have seen, on $\hat{\Sigma}$ this flatness need not be explicitly demanded!} connection $A_\infty$ invariant under $\text{Ad}(G)$ at infinity at the right rate. From now on we actually trivialize $P$, unlike in the previous Sections. This does not mean that our bundle-theoretic treatment so far was purely cosmetic. Rather, we used the formalism of untrivialized principal bundles and affine spaces to develop a trivialization-independent understanding of the allowed gauge transformations in the presence of boundary conditions. Now that we know that this formalism leads us to a choice of Dirichlet boundary condition which all other gauge fields must approach asymptotically, we can simplify our calculations by trivializing the bundle in such a way that our Dirichlet boundary condition appears as the zero gauge field at infinity.

Then $\text{Ad}(P)=P\times_\text{Ad}\mathfrak{g}\cong \Sigma\times\mathfrak{g}$, so that $Q_{A_\infty}\subset\Omega^1(\Sigma,\mathfrak{g})$. To study the momentum map for Yang-Mills theory, we need to know what the phase space looks like. In Section \ref{configspace} we already found the domain of the Lagrangian, namely the tangent bundle to configuration space $TQ\cong Q_{A_\infty}\times Q_\infty$ (the subscript for $Q_\infty$ serves to remind us that these 1-forms vanish asymptotically). The phase space is a dense subspace\footnote{The full cotangent bundle would include distribution-like functionals that are not smooth and which we want to exclude. One could of course also consider restricting $Q_{A_\infty}$ further and allow for the full cotangent bundle $T^*Q$. For instance, one could considering taking $Q$ to consist of only Schwarz functions, so that $T^*Q$ consists of tempered distributions. The power of our argument in this article is that such alterations would not change the main result that the asymptotic symmetry group is the global gauge group.} $\mathcal{P}:=Q_{A_\infty}\times\Omega^2_\infty(\Sigma,\mathfrak{g})\subset T^*Q_{A_\infty}$ of the cotangent bundle \cite{marsdenIntroductionMechanicsSymmetry1999}. It consists of pairs $(A,E)$ with $A\in Q_{A_\infty}$ and $E\in\Omega^2_\infty(\Sigma,\mathfrak{g})\subset T^*_AQ_{A_\infty}$. Here $\Omega^2_\infty(\Sigma,\mathfrak{g})$ denotes the space of 2-forms that approach zero asymptotically at the appropriate rate to be square-integrable. These 2-forms can indeed be viewed as elements of the cotangent space $T_A^*Q_{A_\infty}$, which consists of covectors $T_AQ_{A_\infty}\to\R$, through their action on an element $\alpha_A \in T_AQ_{A_\infty}\cong Q_\infty\subset\Omega^1(\Sigma,\mathfrak{g})$ by means of the conjugate pairing \cite{armsLinearizationStabilityGravitational1979}
\begin{align}\label{canonicalpairing}
    E(\alpha_A)=\langle \alpha_A,E\rangle=\int_\Sigma \text{Tr }\left(\alpha_A\wedge E\right).
\end{align}
The constraint for Yang-Mills theory is the Gauss law \cite{armsStructureSolutionSet1981}
\begin{align*}
    D_AE:=dE+[A\wedge E]=0\,.
\end{align*}
The action of the boundary-preserving gauge group $\mathcal{G}^I$ lifts to phase space in the obvious way:
\begin{align*}
\forall g\in\mathcal{G}^I:~~~~~~g\cdot(A,E)=(g^{-1}Ag+g^{-1}dg,g^{-1}Eg)\,.
\end{align*}
The Lie algebra $\text{Lie}(\mathcal{G}^I)$ is isomorphic to  $C_I^\infty(\Sigma,\mathfrak{g})$, i.e. the space of smooth gauge transformation parameters that leave the boundary conditions invariant by becoming constant towards infinity at the right rate (we recall that this rate is determined by the conformal factor from the previous Section). We equip $\mathcal{P}\subset T^*Q_{A_\infty}$ with the canonical symplectic form $\omega=\int_\Sigma\text{Tr }\mathbbm{d}A\wedge\mathbbm{d}E$, where the $\mathbbm{d}$ symbol is used to stress that this is the derivative operator on the infinite-dimensional phase space of fields and not the $d$ on 3-space $\Sigma$. Henceforth we will occasionally use double-slashed symbols to stress that these objects are defined on infinite-dimensional phase space $\mathcal{P}$.

We should like to check that, with this symplectic form, the Gauss law constraint generates gauge transformations, i.e. check for which gauge parameters $\xi\in\text{Lie}(\mathcal{G}^I)$ the momentum map equals the Gauss law. These calculations are not novel, in that they largely establish a well-known fact about the momentum map for Yang-Mills theory, but they are nonetheless a necessary pedagogical review for our subsequent discussion of the precise rates at which gauge transformations generated by the Gauss constraint must approach the identity asymptotically.

Let us recall the definition of the momentum map \cite{dasilvaLecturesSymplecticGeometry2008}.

\begin{definition}\label{mommap}
Let $(\mathcal{P},\omega)$ be a symplectic manifold and $H$ a Lie group that acts on $\mathcal{P}$ by symplectomorphisms.\footnote{I.e.~the action of $H$ preserves the symplectic form $\omega$.} Let $\mathfrak{h}$ denote the Lie algebra of $H$ with dual $\mathfrak{h}^*$, and write $\langle\cdot,\cdot\rangle\colon \mathfrak{h}^*\times\mathfrak{h}\to\R$ for the pairing of the algebra and its dual. Then a \textit{momentum map} for the $H$-action on $\mathcal{P}$ is an equivariant\footnote{With respect to the $H$-action on $\mathcal{P}$ and the coadjoint action on $\mathfrak{h}^*$.} map $\mu\colon \mathcal{P}\to\mathfrak{h}^*$ such that, for all $\xi\in\mathfrak{h}$, we have:
\begin{align*}
\mathbbm{d}\langle\mu,\xi\rangle=\iota_{\mathbb{X}_\xi}\omega=\omega(\mathbb{X}_\xi,\cdot)\,.
\end{align*}
Here, $\mathbb{X}_\xi$ denotes the fundamental vector field\footnote{In this definition we use the double-slashed notation because this agrees with our subsequent calculations, but of course this definition of the momentum map is also valid for finite-dimensional symplectic manifolds.} generated by $\xi$, and $\langle\mu,\xi\rangle$ is understood as a function $\langle\mu,\xi\rangle\colon \mathcal{P}\to\R$, defined as follows: $\langle\mu,\xi\rangle(x)=\langle\mu(x),\xi\rangle$.
\end{definition}

The idea behind this definition is that the fundamental vector field $\mathbb{X}_\xi$ infinitesimally generates the $H$-action with parameter $\xi$, while the values $\langle\mu,\xi\rangle$ of the momentum map for specific $\xi$ provide constants of motion. The required relation $\mathbbm{d}\langle\mu,\xi\rangle=\omega(\mathbb{X}_\xi,\cdot)$ can then be viewed in the light of Noether's theorem: it relates the conservation of the constants of motion to the symmetry of the theory.\footnote{For technical details see \cite{marsdenIntroductionMechanicsSymmetry1999,dasilvaLecturesSymplecticGeometry2008,binz}, for a conceptual exposition see \cite{Butterfield2006}.} For gauge symmetries it is a version of the generation of gauge symmetries by taking Poisson brackets of fields with the smeared Gauss law.

We will now check that the Gauss law constraint is indeed the momentum map for the gauge group. We find that, by partial integration, the smeared Gauss constraint splits into a ``bulk'' term corresponding to the infinitesimally generated gauge symmetries and a boundary term (the electric flux). The boundary term must vanish, leading to a condition on the gauge transformation parameters. We will present our derivation on the Cauchy surface $\Sigma\cong \R^3$, interpreting $\partial\Sigma$ as an asymptotic boundary, but the exact same derivation would work on the compact space $\hat{\Sigma}$ with actual boundary $\partial\hat{\Sigma}$, as long as one takes care of the appropriate conformal factors that appear in the integrals. We will comment more on the conformal behavior of the electric field and the electric flux in Section \ref{electricfluxsection}.

The momentum map $\mu\colon \mathcal{P}\to \Omega_I^3(\Sigma,\mathfrak{g})\subset \text{Lie}(\mathcal{G}^I)^*$, where the $I$ denotes appropriate asymptotic fall-off behavior, for the action of the gauge group $\mathcal{G}^I$ on $\mathcal{P}\subset T^*Q_{A_\infty}$ is supposed to be the Gauss law\footnote{If we consider Maxwell theory, then the momentum map $\mu$ applied to an element $\xi\in C_I^\infty(\Sigma,\mathfrak{g})$ is just the familiar Gauss law $\nabla\cdot\textbf{E}$ smeared with $\xi$. This can be seen by switching to the physicists' convention $\xi=i\lambda$ and writing $D_AE=\nabla\cdot\textbf{E}$, yielding $i\int_\Sigma d^3x~ \lambda(\mathbf{x})\,\nabla\cdot\textbf{E}(\mathbf{x})$.} constraint $\mu(A,E)=D_AE$ \cite{armsStructureSolutionSet1981,binz}.
Here we identify $\eta\in\Omega_I^3(\Sigma,\mathfrak{g})$ as an element in the dual $C^\infty_I(\Sigma,\mathfrak{g})^*$, through the pairing $\langle \eta,\xi\rangle=\int_\Sigma \text{Tr}\left(\xi\wedge\eta\right)$, similar to the pairing defined earlier. 

\begin{proposition}
    If one requires $E\to 0+\mathcal{O}(r^{-2})$, then gauge parameters $\xi$ need to approach zero without any particular fall-off rate in order for the Gauss constraint to be the momentum map. If instead we demand $E\to 0+\mathcal{O}(r^{-3/2-\epsilon})$, then we must have $\xi\to 0+\mathcal{O}(r^{-1/2})$.
\end{proposition}

\noindent\textit{Proof.} For any $\xi\in C_I^\infty(\Sigma,\mathfrak{g})$ (and using Stokes' theorem/partial integration), we have:
\begin{align}\label{momentummappartial}
\begin{split}
    &\langle\mu,\xi\rangle(A,E)=\int_\Sigma\text{Tr }D_A E\wedge\xi=\int_\Sigma\text{Tr }(dE+[A,E])\wedge\xi=\int_\Sigma\text{Tr}\left(dE\wedge\xi-[E,A]\wedge\xi\right)
    \\
    &=-\int_\Sigma\text{Tr }E\wedge d\xi+\int_{\partial \Sigma}\text{Tr }E\wedge\xi-\int_\Sigma\text{Tr }E\wedge[A,\xi]=-\int_\Sigma\text{Tr }E\wedge D_A\xi+\int_{\partial \Sigma}\text{Tr }E\wedge\xi\,,
\end{split}
\end{align}
where we have used the ad-invariance of the trace, i.e. $\text{Tr}\left([E,A]\wedge \xi\right)=\text{Tr}\left(E\wedge[A,\xi]\right)$. Note that for consistency we have used the $\wedge$ symbol even on $\xi\in C^\infty_I(\Sigma,\mathfrak{g})$, even though it is a 0-form.

But, if $\mu(A,E)=D_AE$ really is to define the momentum map for the action of $\mathcal{G}^I$, then by definition it must satisfy the property
\begin{align}\label{momentummaprequirement}
\mathbbm{d}\langle\mu,\xi\rangle=\iota_{\mathbb{X}_\xi}\omega:=\omega(\mathbb{X}_\xi,\cdot),\;\;\;\;\; \xi\in C_I^\infty(\Sigma,\mathfrak{g}),
\end{align}
where $\mathbb{X}_\xi\in\mathfrak{X}(\mathcal{P})$ denotes the fundamental vector field on $\mathcal{P}$ generated by the Lie algebra element $\xi$. We will now check what assumption on the asymptotic behavior of the gauge transformation parameter is required for the above condition to hold.

To this end we first calculate the right- and left-hand sides of Eq.~\eqref{momentummaprequirement} separately and then compare them. We begin with the right-hand side, i.e.~$\omega(\mathbb{X}_\xi,\cdot)$. By definition, for any function $\mathbb{F}\in C^\infty(\mathcal{P})$, we have:
\begin{align*}
    \mathbb{X}_\xi(\mathbb{F})(A,E)=\frac{d}{dt}\bigg|_{t=0}\mathbb{F}\left(e^{t\xi}\cdot (A,E)\right)=\frac{d}{dt}\bigg|_{t=0}\mathbb{F}\left(e^{-t\xi}Ae^{t\xi}+e^{-t\xi}d(e^{t\xi}),e^{-t\xi}Ee^{t\xi}\right).
\end{align*}
For the functions $\mathbb{F}=A$ and $E$, this simply gives:
\begin{align*}
    \mathbb{X}_\xi(A)&=\frac{d}{dt}\bigg|_{t=0}\left(e^{-t\xi}Ae^{t\xi}+e^{-t\xi}d(e^{t\xi})\right)=-\xi A+A\xi+d\xi=[A,\xi]+d\xi=D_A\xi\,,
    \\
    \mathbb{X}_\xi(E)&=\frac{d}{dt}\bigg|_{t=0}\left(e^{-t\xi}Ee^{t\xi}\right)=-\xi E+E\xi=[E,\xi]\,.
\end{align*}
Thus if we put $\mathbb{X}_\xi$ in the first slot of the symplectic form $\omega=\int_\Sigma\text{Tr }\mathbbm{d}A\wedge\mathbbm{d}E$, i.e.~the right-hand side of Eq.~\eqref{momentummaprequirement}, we get:
\begin{align}\label{RHS}
\begin{split}
    \omega_{(A,E)}(\mathbb{X}_\xi,\cdot)&=\int_\Sigma\text{Tr }  \left(\mathbbm{d}A(\mathbb{X}_\xi)\wedge\mathbbm{d}E-\mathbbm{d}E(\mathbb{X}_\xi)\wedge\mathbbm{d}A\right)=\int_\Sigma\text{Tr } \left(\mathbb{X}_\xi(A)\wedge\mathbbm{d}E-\mathbb{X}_\xi(E)\wedge\mathbbm{d}A\right)
    \\
    &=\int_\Sigma\text{Tr } \left(([A,\xi]+d\xi)\wedge\mathbbm{d}E-[E,\xi]\wedge\mathbbm{d}A\right)=\int_\Sigma\text{Tr }(D_A\xi\wedge\mathbbm{d}E-[E,\xi]\wedge\mathbbm{d}A)\,.
\end{split}
\end{align}
The left hand-side of Eq.~\eqref{momentummaprequirement} gives:
\begin{align*}
    \mathbbm{d}\langle\mu,\xi\rangle=\mathbbm{d}\int_\Sigma\text{Tr } D_AE\wedge\xi=\int_\Sigma\text{Tr }\left(\mathbbm{d}(D_AE)\wedge \xi-D_AE\wedge\mathbbm{d}\xi\right).
\end{align*}
However, we cannot immediately see how this agrees with the expression in Eq.~\eqref{RHS}, because Eq.~$\eqref{RHS}$ contains a term linear in $D_A\xi$, while the above result has a term that is linear in $\xi$. Thus we need to do the partial integration in Eq.~\eqref{momentummappartial}, which gives:
\begin{align}\label{LHSpartial}
    \mathbbm{d}\langle\mu,\xi\rangle=-\int_{\Sigma}\text{Tr }(\mathbbm{d}E\wedge D_A\xi-E\wedge\mathbbm{d}(D_A\xi))+\mathbbm{d}\int_{\partial \Sigma}\text{Tr }E\wedge\xi\,.
\end{align}
The second term in the first integral can be rewritten as:
\begin{align*}
    E\wedge\mathbbm{d}(D_A\xi)&=E\wedge\mathbbm{d}(d\xi+[A,\xi])=E\wedge\mathbbm{d}[A,\xi]=E\wedge[\mathbbm{d}A,\xi]=E\wedge \mathbbm{d}A\xi-E\wedge\xi\mathbbm{d}A
    \\
    &=-E\xi\wedge\mathbbm{d}A+\xi E\wedge\mathbbm{d}A-\xi E\wedge\mathbbm{d}A+E\wedge \mathbbm{d}A\xi=-[E,\xi]\wedge\mathbbm{d}A+[E\wedge\mathbbm{d}A,\xi]\,.
\end{align*}
Thus, the first integral in Eq.~\eqref{LHSpartial} equals:
\begin{align*}
    \int_\Sigma\text{Tr }(D_A\xi\wedge\mathbbm{d}E+E\wedge\mathbbm{d}(D_A\xi))=\int_\Sigma\text{Tr }(D_A\xi\wedge\mathbbm{d}E-[E,\xi]\wedge\mathbbm{d}A+[E\wedge\mathbbm{d}A,\xi])\,.
\end{align*}
But the trace of the full commutator term gives zero,\footnote{Or, since the trace is ad-invariant, we could also immediately have rewritten $\text{Tr }E\wedge[\mathbbm{d}A,\xi]=-\text{Tr }[E,\xi]\wedge\mathbbm{d}A$.}
so we obtain precisely the final expression in Eq.~\eqref{RHS}! This implies that from requiring that the Gauss constraint $\mu(A,E)=D_AE$ is the momentum map for the action of the gauge group, it follows that the boundary term in Eq.~\eqref{LHSpartial} must be zero. Since that boundary term is an integral of $E\wedge\xi$, the condition that needs to be imposed on $\xi$ to ensure that this term vanishes depends on the asymptotic fall-off behavior of $E$. If we take $E\to 0+\mathcal{O}(r^{-2})$, then $\xi$ just needs to go to zero without any particular fall-off rate (since the integral is over a 2-sphere with radius sent to infinity, which grows as $r^2$). If instead we demand $E\to 0+\mathcal{O}(r^{-3/2-\epsilon})$, then we must have $\xi\to 0+\mathcal{O}(r^{-1/2})$ to guarantee that there is no boundary term. $\Box$
\\
\\
\indent However, if we require only slightly stronger asymptotic fall-off conditions on the electric field, e.g. $E\to 0+\mathcal{O}(r^{-2-\epsilon})$, then apparently there no longer is any need for asymptotic requirements on $\xi$. Thus this approach does not provide a completely unambiguous and satisfactory answer to the question of precisely what asymptotic behavior of gauge transformations is required to be able to call them redundant. Moreover, even if we do conclude that we must have e.g. $\xi\to 0+\mathcal{O}(r^{-1/2})$, it is not yet clear that the quotient $\mathcal{G}^I/\mathcal{G}^\infty_0$ will be precisely the group of global gauge transformations, even though this global group can be pristinely deduced from other approaches, as was explained in Section \ref{intro}. We will definitively resolve this issue in Section \ref{localizable}.

\begin{remark}
    In our framework, the fall-off conditions ensuring existence of the instantaneous Lagrangian do not define a second constraint imposed on top of a larger unconstrained phase space. Rather, they define the function spaces that constitute the phase space $\mathcal{P}:=Q_{A_\infty}\times\Omega^2_\infty(\Sigma,\mathfrak{g})$ itself where the square-integrability of fields is built into the definitions of $Q_{A_\infty}$ and $\Omega^2_\infty(\Sigma,\mathfrak{g})$ from the outset. This is necessary in particular to ensure that the canonical symplectic form $\omega=\int_\Sigma\text{Tr }\mathbbm{d}A\wedge\mathbbm{d}E$ is well-defined. The constraint surface $\mathcal{C}$ is then defined within $\mathcal{P}$ solely by the Gauss constraint, i.e.
    \begin{align*}
        \mathcal{C}=\mu^{-1}(0).
    \end{align*}
    In this framework, ``allowed'' gauge transformations are simply those whose action on instantaneous state space as well as phase space is well-defined, ensuring also that the presymplectic form $\omega|_\mathcal{C}$ is inherited in a well-defined way. Thus, in this respect, the situation is not essentially different from a system without boundary conditions. In particular, the constraint surface is known to be coisotropic if 0 is a regular value of the momentum map, as is the case in electromagnetism. In non-Abelian Yang-Mills theory, however, the momentum map is singular and only regular at the irreducible connections, on which the gauge group acts freely with trivial stabilizer \cite{singerGeometryOrbitSpace1981,Mitter:1979un,Kondracki1986Stratification,ARMS199043}. Thus the failure of the constraint surface to be coisotropic at singular points is a general feature of non-Abelian Yang-Mills theory, regardless of whether we are working on a space with or without boundaries. In this sense the coisotropic/singular nature of the constraint surface is an independent question from the one we are aiming to answer in this article, as we are aiming to find the asymptotic fall-off conditions required to yield a well-defined instantaneous classical field theory in the first place.
\end{remark}

\subsection{Electric flux through infinity}\label{electricfluxsection}

Before we move on we resolve a paradox concerning the case in which we require asymptotic fall-off behavior of order $\mathcal{O}(r^{-2})$ on both the gauge and electric fields. As we saw in Section \ref{conformal}, this specific behavior can be formalized by means of a conformal compactification of Minkowski spacetime with conformal factor $K\sim r^{-2}$. But how does the electric field $E$ transform under this conformal compactification? At the beginning of this Section, we defined electric fields in the Hamiltonian picture as 2-forms in $\Omega^2_\infty(\Sigma,\mathfrak{g})\subset T^*Q_{A_\infty}$ which act on tangent vectors $\alpha_A\in T_AQ_{A_\infty}$ through the canonical pairing in Eq. \eqref{canonicalpairing}. Since 2-forms are not defined with reference to the metric, electric fields $E$ do not transform with any conformal factor under the conformal embedding $\Sigma\to \hat{\Sigma}$ \cite{SNIATYCKI1988291}. Thus it seems that, similar to the ``electric fields'' $\alpha_A$, the asymptotic fall-off requirement $E\to 0+\mathcal{O}(r^{-2})$ simply translates to the requirement that $\hat{E}$ vanishes on $\partial\hat{\Sigma}$. But this raises a paradox: the integral $\int_\Sigma\text{Tr}\left(D_A E\right)$
equals the electric flux, which can be nonzero since by Gauss's theorem this integral is just a limit of an integral over a sphere of radius $r$ with $r\to\infty$. This integral grows with $r^2$, canceling the fall-off behavior $\mathcal{O}(r^{-2})$. On the other hand, if $\hat{E}$ is zero on $\partial\hat{\Sigma}$, then the flux
\begin{align*}
    \int_{\hat{\Sigma}}\text{Tr}\left(D_{\hat{A}}\hat{E}\right)=\int_{\partial\hat{\Sigma}}\text{Tr}\left(\hat{E}\right)
\end{align*}
vanishes. But the two calculations of the electric flux are supposed to agree.

The problem with this line of thinking lies in the precise definition of fall-off behavior for forms. The notation $\mathcal{O}(r^{-2})$ only makes sense for functions, which can be obtained from forms by feeding them vector fields. But if $E$ is a 2-form, then it needs to be fed two vector fields to produce a function, so in coordinates it has two indices. This does not coincide with our intuition of the electric field as a vector field $\textbf{E}$ with three components $E^i$, each of which must fall off asymptotically with order at least $r^{-3/2-\epsilon}$ to ensure that $\norm{\textbf{E}}^2$ is integrable. Instead, it is more sensible to use the Hodge star involution to write $E=*\mathcal{E}$, where $\mathcal{E}\in\Omega^1(\Sigma,\mathfrak{g})$ is a 1-form, and to require $\mathcal{E}_i\to 0+\mathcal{O}(r^{-2})$. The Gauss law then becomes $D_A*\mathcal{E}=0$ \cite{SNIATYCKI1988291}. The 1-form $\mathcal{E}$ exhibits the exact same conformal behavior as $\alpha_A$, meaning that $\hat{\mathcal{E}}$ must vanish on $\partial\hat{\Sigma}$ to ensure the square-integrability of its pullback to $\Sigma$. The 2-form $E=*\mathcal{E}$, however, is defined through the Hodge star in terms of the metric, and therefore it does pick up\footnote{To see this, consider the simpler example of the function $1$ which is identically 1 on $\Sigma$. This function does not pick up any conformal factor. But $d\text{Vol}=*1$ does, since in coordinates it is the square root of the determinant of the metric.} a conformal factor: $\hat{E}=K^{-1}E$ \cite{SNIATYCKI1988291}. This ensures that $\hat{E}$ can be any smooth function on $\hat{\Sigma}$ (not necessarily vanishing on the boundary), so that there can indeed be a nonzero electric flux through infinity.

With this complication clarified, there is a useful conclusion to be drawn from the derivation in Section \ref{mommapsection}. By partial integration the momentum map naturally falls into two parts, i.e. two integrals, viz. $\int_\Sigma\text{Tr }E\wedge D_A\xi$ and the boundary term $\int_{\partial \Sigma}\text{Tr }E\wedge\xi$. The first corresponds precisely to the symmetries that are infinitesimally generated by the fundamental vector fields $\mathbb{X}_\xi$, whereas the second does not \cite{Tanzi:2021xva} - it equals the smeared electric flux through infinity \cite{rielloHamiltonianGaugeTheory2024}. Symmetries generated by the fundamental vector fields $\mathbb{X}_\xi$ therefore satisfy the requirement that $\xi$ vanishes asymptotically, i.e. that $\hat{\xi}|_{\partial\hat{\Sigma}}=0$. This means that only gauge transformations vanishing at infinity, i.e. local transformations, are associated to the Gauss law constraint through Noether's second theorem \cite[Proposition 7.2.6]{binz}. Global gauge transformations, which do act non-trivially at infinity, are not included and only appear in Noether's first theorem \cite{Noether1918}.

\subsection{Infinitesimal localizable symmetries}\label{localizable}

In the conformal picture, the boundary-preserving gauge transformations are constant on $\partial\hat{\Sigma}$ and the unphysical transformations vanish on $\partial\hat{\Sigma}$. The physical gauge group therefore equals a copy of $G$ on the boundary, viewed as the global gauge group of constant transformations. But a weakness of this conclusion is that it seems to depend on the choice of asymptotic boundary conditions for the fields. If we really want the weakest possible conditions that still ensure a finite Lagrangian, i.e. fall-off with order $\mathcal{O}(r^{-3/2-\epsilon})$ on $\alpha_A$ (and therefore $A$ itself) and on $\mathcal{E}=*E$, then the conformal analysis is inapplicable because the conformal factor $K\sim r^{-2}$ suppresses the radial components of the 1-forms too strongly. This issue can be remedied, however, by means of the following result.
\begin{proposition}\label{falloffprop}
Gauge parameters $\xi$ which preserve $\mathcal{O}(r^{-3/2-\epsilon})$ fall-off, i.e. which are such that $\partial_i\xi\to 0+\mathcal{O}(r^{-3/2-\epsilon})$, must satisfy $\xi\to\text{const}+\mathcal{O}(r^{-1/2-\epsilon})$ (although the converse does not hold, think e.g. of $\sin(r)/r$). 
\end{proposition}

\noindent\textit{Proof.} Taking $\xi$ to be real-valued for simplicity (but with obvious generalization to cases other than $G=U(1)$), by the fundamental theorem of calculus:
\begin{align*}
    \left|\lim_{s\to\infty}\xi(s,\theta,\phi)-\xi(r,\theta,\phi)\right|=\left|\int_r^\infty\partial_s\xi(s,\theta,\phi)ds\right|\leq \int_r^\infty Cs^{-3/2-\epsilon}ds\leq 2Cr^{-1/2-\epsilon}.
\end{align*}
From this it follows that $\lim_{s\to\infty}\xi(s,\theta,\phi)=:c_{\theta,\phi}$ exists and that $|\xi(r,\theta,\phi)-c_{\theta,\phi}|=\mathcal{O}(r^{-1/2-\epsilon})$. A priori it could be the case that the constant $c_{\theta,\phi}$ is different in each angular direction, but it is straightforward to show that this cannot be so. At fixed $r$ we can connect any two points $x,y$ by a curve $\gamma$ of length at most $\pi r$. This factor of $r$ is cancelled by the $1/r$ that appears in the angular components of the gradient $\nabla\xi$ expressed in spherical coordinates. Integrating over $\gamma$ we find that $|\xi(x)-\xi(y)|\leq \pi\tilde{C}r^{-1/2-\epsilon}$. This shows that $c_{\theta,\phi}=c_{\theta',\phi'}$. Thus we find that any gauge parameter $\xi$ whose derivative $d\xi$ is square-integrable satisfies $\xi\to\text{const}+\mathcal{O}(r^{-1/2-\epsilon})$. $\Box$
\\
\\
\indent But the unphysical gauge transformations generated by the Gauss constraint are precisely the subset of these gauge parameters for which the constant at infinity equals zero. Thus the quotient of these two algebras of gauge transformations is just $\mathfrak{g}=\text{Lie}(G)$, which when exponentiated gives the global gauge group.

If, on the other hand, we choose stronger conditions, e.g. $E\to 0+\mathcal{O}(r^{-3})$ or that $E$ is a Schwarz function, then the boundary term in Eq. \eqref{momentummappartial} automatically vanishes, regardless of the asymptotic behavior of the gauge parameter $\xi$, obviating the need for the condition $\xi|_{\partial\hat{\Sigma}}=0$. The goal of the remainder of this Section is to explain why global gauge symmetries nonetheless \textit{never} play a role in Noether's second theorem, i.e. why they do not give rise to constraints, and should therefore not be considered to be redundant even if the boundary term in Eq. \eqref{momentummappartial} vanishes due to stricter boundary conditions.

 In the mathematical physics literature the symmetries that give rise to constraints through Noether's second theorem are the so-called \textit{infinitesimal localizable symmetries}. These form an ideal (under the Lie bracket) $\mathfrak{G}\subset\text{Lie}(\mathcal{G}^I)$ of the Lie algebra of the full boundary-preserving symmetry group, and the constraint surface is the zero locus of the momentum map for the infinitesimal localizable symmetries (see Section 7.5 of \cite{binz} or \cite{sniatycskiboundarycondspatboundeddom,rielloHamiltonianGaugeTheory2024,Gotay2006MomentumMA}).
For this reason the infinitesimal localizable symmetries should be identified as the redundant, ``trivial'' ones. When exponentiated, they generate the minimal symmetry group that must be called \textit{gauge} in the sense of ``unphysical'' in order to guarantee an appropriate form of determinism. These infinitesimal localizable symmetries are introduced in Definition 7.2.5 of \cite{binz}, but we will not reproduce that definition here, since it is based on the formalism of jet bundles. However, when adapted to our case at hand, it reads as follows \cite{sniatycskiboundarycondspatboundeddom,Gotay2006MomentumMA}:
\begin{definition}
    An infinitesimal symmetry $\xi\in\text{Lie}(\mathcal{G}^I)=C^\infty_I(\Sigma,\mathfrak{g})$ is called \textit{localizable} if it vanishes on the (asymptotic) boundary of $\Sigma$ and if for any pair of open sets $U,V\subset \Sigma$ with disjoint closures, there exists a $\xi'\in\text{Lie}(\mathcal{G}^I)$ such that 
    \begin{align*}
        &\xi(x)=\xi'(x),\;\;\;\;\; x\in U;
        \\
        &\xi'(x)=0,\;\;\;\;\;\;\;\;\;\; x\in V.
    \end{align*}
\end{definition}
In other words: an infinitesimal symmetry is localizable if it is zero at asymptotic infinity and for any two disjoint open regions we can always find another infinitesimal symmetry that is equal to the original one on the one region, but zero on the other. That is: we can always localize the infinitesimal symmetry to some open region of space.

Clearly, global gauge transformations are not localizable since they do not vanish at asymptotic infinity, or more precisely, at the boundary $\partial\hat{\Sigma}$ of the compactified space $\hat{\Sigma}$ from Section \ref{boundarycond}. The question, then, is whether \textit{all other} infinitesimal symmetries in $\text{Lie}(\mathcal{G}^I)$ are localizable. If this is so, then the quotient $\mathcal{G}_0^I/\mathcal{G}^\infty_0$ equals precisely the global gauge group $G$, where $\mathcal{G}^I_0$ is the identity component of $\mathcal{G}^I$ and $\mathcal{G}^\infty_0$ denotes the group generated by all $e^\xi$ with $\xi\in\mathfrak{G}$.

\begin{proposition}
All gauge symmetries except the global ones are localizable.
\end{proposition}

\noindent\textit{Proof.} This is done most easily by working on $\hat{\Sigma}$. There $\text{Lie}(\mathcal{G}^I)$ consists of all maps $\hat{\xi}\colon \hat{\Sigma}\to\mathfrak{g}$ that are constant on $\partial\hat{\Sigma}$. We note that $\text{Lie}(\mathcal{G}^I)/\mathfrak{g}$, with $\mathfrak{g}$ viewed as the constant maps in $C^\infty_I(\Sigma,\mathfrak{g})$, is isomorphic to the algebra $\mathfrak{G}_\infty$ of all maps $\hat{\xi}\colon \hat{\Sigma}\to\mathfrak{g}$ that vanish on $\partial\hat{\Sigma}$. We need to show that $\mathfrak{G}_\infty=\mathfrak{G}$, i.e. that the algebra of infinitesimal localizable symmetry consists precisely of the gauge parameters that vanish on $\partial\hat{\Sigma}$.

There are two situations to consider: if $U,V$ are the open subsets from the above definition, such that $\hat{\xi}\in\text{Lie}(\mathcal{G}^I)$ must be localized on $U$ relative to $V$, then either $U$ could lie in the interior of $\hat{\Sigma}$ or contain (part of) the boundary $\partial\hat{\Sigma}$. In the first case it is obvious that $\hat{\xi}$ can be localized: we just use a $\mathfrak{g}$-valued bump function $\hat{f}$ that is the identity on $U$ and becomes zero very quickly outside of $U$, in particular on $V$. It is then clear that $\hat{f}\cdot\hat{\xi}$ will be the required element of $\text{Lie}(\mathcal{G}^I)$ that agrees with $\hat{\xi}$ on $U$ and is zero on $V$. In the case in which $U$ contains part of the boundary it is not immediately clear whether $\hat{f}\cdot\hat{\xi}\in\text{Lie}(\mathcal{G}^I)$. But since $\hat{\xi}$ is zero on $\partial\hat{\Sigma}$, so is the product $\hat{f}\cdot\hat{\xi}$. This means $\hat{f}\cdot\hat{\xi}\in\mathfrak{G}_\infty\subset\text{Lie}(\mathcal{G}^I)$. We conclude that the algebra of infinitesimal localizable symmetries $\mathfrak{G}$ is indeed $\mathfrak{G}_\infty$. $\Box$
\\
\\
\indent Thus we find that, in Yang-Mills theory, localizability effectively reduces to just the condition of vanishing at infinity.\footnote{Note that this result is quite independent of the precise form of $Q$ and $\mathcal{G}^I$. No matter what asymptotic conditions on the fields are required, we always find that the redundant gauge transformations are all elements of $\mathcal{G}^I$ which vanish at infinity.} Of course, this is not a surprising result, since gauge symmetries are meant to be localizable. But we clearly see that if a field theory contains only global (rigid) symmetries, then no infinitesimal transformation is ever localizable, in which case $\mathfrak{G}$ would be zero and there would be no constraints.

This means that as long as the asymptotic boundary can be formalized as a conformal boundary, the subalgebra $\mathfrak{G}_\infty$ generates (through the exponential map) the subgroup $\mathcal{G}^\infty_0\subset\mathcal{G}^I$ of gauge transformations that become the identity at asymptotic infinity at the appropriate rate (which equals the rate at which elements of $\mathcal{G}^I$ approach a constant) and lie in the identity component of $\mathcal{G}^I$ (i.e. can be obtained by exponentiating Lie algebra elements). The quotient of physical gauge transformations
\begin{align*}
    \mathcal{G}_\text{Phys}=\mathcal{G}^I/\mathcal{G}^\infty_0
\end{align*}
then looks like a copy of the global gauge group $G$, corresponding to all possible constants at infinity, for every homotopy class of gauge transformations, i.e. for every connected component of the gauge group $\mathcal{G}^I$. In three dimensions these homotopy classes are determined by the fundamental group $\pi_3(G)$, since gauge transformations on $\Sigma$ that are constant at asymptotic infinity can be viewed as maps $S^3\to G$, where asymptotic infinity corresponds to one point on $S^3$, e.g. the North pole. For $G=U(1)$ this homotopy group is trivial,\footnote{In one dimension we do have interesting topology for electromagnetism since $\pi_1(S^1)\cong\Z$.} but for $G=SU(2)$ we have $\pi_3(SU(2))\cong\pi_3(S^3)\cong\mathbb{Z}$.

Since by Proposition \ref{falloffprop}, and the argument right below its proof, the quotient $\mathcal{G}^I/\mathcal{G}^\infty_0$ still equals (several homotopy copies of) the global gauge group when $\mathcal{O}(r^{-3/2-\epsilon})$ fall-off is imposed on the electric fields, we have covered all major scenarios and obtained the same result for $\mathcal{G}_\text{Phys}$ in each.

\section{Adding the Higgs field}\label{higgs}

Over the past two decades there has been a substantial conceptual debate about the Higgs mechanism \cite{earmanCuriePrincipleSpontaneous2004,smeenkElusiveHiggsMechanism2006,lyreDoesHiggsMechanism2008,struyveGaugeInvariantAccounts2011a,stoltznerConstrainingHiggsMechanism2012,fraserHiggsMechanismSuperconductivity2016}. Much of this debate centers around the physical status of gauge symmetries in relation to gauge symmetry breaking. As has been pointed out in \cite{lusannaDiracObservablesHiggs1997,lusannaDiracObservablesHiggs1997a,struyveGaugeInvariantAccounts2011a,wallaceIsolatedSystemstwee}, a key point is that the unbroken and broken phases of the Higgs model exhibit differing asymptotic boundary conditions. However, the derivation of this point has not been performed rigorously. We can now do this in the framework developed in the previous Sections.

To include a Higgs field, we must enlarge the configuration space $Q_{A_\infty}$ of Yang-Mills fields to $Q_{A_\infty}\times \Tilde{Q}$, where $\Tilde{Q}$ is the space of Higgs fields, which are sections of an associated vector bundle $P\times_\rho V\to \Sigma$ through a representation $\rho\colon G\to\text{GL}(V)$, where $V$ is the Higgs vector space.\footnote{It is $\C$ for $G=U(1)$, and $\C^2$ for both $G=SU(2)$ and $G=U(1)\times SU(2)$ \cite{hamiltonMathematicalGaugeTheory2017}.} If we equip $V$ with an inner product $\langle\cdot,\cdot\rangle$, then we can define a norm on the sections in $\Gamma(P\times_\rho V)$ in a similar way as for the gauge fields: by integrating the absolute value of such a section over $\Sigma$.

The tangent space $T_\varphi \Tilde{Q}$ at a point $\varphi\in\Gamma(P\times_\rho V)$ is itself just isomorphic to $\Tilde{Q}$. However, we need to restrict both $\Tilde{Q}$ and $T\Tilde{Q}$ with appropriate asymptotic boundary conditions. These follow from the instantaneous Yang-Mills-Higgs Lagrangian,\footnote{For the well-posedness of the Yang-Mills-Higgs initial value problem see \cite{eardleyGlobalExistenceYangMillsHiggs1982}.} which is given by
\begin{align*}
    \mathcal{L}_\text{YMH}(A,\alpha_A,\varphi,\psi_\varphi)=\frac{1}{2}\norm{\alpha_A}^2-\frac{1}{2}\norm{F(A)}^2+\frac{1}{2}\norm{\psi_\varphi}^2-\frac{1}{2}\norm{D_A\varphi}^2-\int_\Sigma V(\varphi) d\text{Vol},
\end{align*}
for $A\in Q,\alpha_A\in T_AQ, \varphi\in\Tilde{Q},\psi_\varphi\in T_\varphi\Tilde{Q}$. Here $V(\varphi)$ is the well-known Higgs potential, $D_A\varphi$ is the covariant derivative of the Higgs field and $\psi_\varphi\in T_\varphi\Tilde{Q}$ is thought of as the velocity of $\varphi\in\Tilde{Q}$.

In order to guarantee existence, we require that each individual term in the above instantaneous Lagrangian is finite. We already know that this requires $\alpha_A\to 0$ and $F(A)\to 0$, but now we also need $\psi_\varphi\to 0$ and $D_A\varphi\to 0$, as well as a condition related to $V(\varphi)$. This last condition is ambiguous. If the Higgs potential has the familiar shape
\begin{align*}
    V(\varphi)=-\mu\norm{\varphi}^2+\lambda\norm{\varphi}^4,
\end{align*}
then clearly the zero-point of $V(\varphi)$ lies at $\varphi=0$ if $\mu>0$ (besides some other manifold of roots at which $\varphi\neq 0$). Thus, we expect the boundary condition $\varphi\to 0$. However, we may instead want to think of the \textit{minimum} of $V(\varphi)$ as the true vacuum, therefore requiring $\varphi\to \text{min}$ instead. These two possibilities respectively correspond to the so-called unbroken and broken phases of the Higgs model. Let us now study what the group $\mathcal{G}^I$ of boundary-preserving gauge symmetries looks like in both cases.\footnote{We note that our ideas agree with \cite{wallaceIsolatedSystemstwee}, but fill in the missing argument, namely the Lagrangian must be defined on the tangent bundle to configuration space. Without this added argument one \textit{cannot} deduce that the physical gauge group is different in the two cases.}

\textit{The unbroken phase.} In the unbroken phase, we assume that $\varphi=0$ is the vacuum for the Higgs field, i.e. that this state carries zero energy. We can either think of this state as lying in the symmetric middle of the ``Mexican hat potential,'' or as the potential itself being such that it only has a minimum at $\varphi=0$, e.g. by taking $\mu<0$. Since $\varphi=0$ corresponds to zero energy, we require the asymptotic boundary condition $\varphi\to 0$, besides the common boundary conditions $\psi_\varphi\to 0$ and $D_A\varphi\to 0$ which are always needed. Note that this indeed gives the configuration space at infinity the right structure: the space of Higgs fields at infinity is zero-dimensional, since it consists only of $\varphi=0$. The tangent space at infinity then also consists only of zero, which is what we want since we require $\psi_\varphi\to 0$. Now, the conditions $\psi_\varphi\to 0$ and $D_A\varphi\to 0$ are always preserved by any gauge transformation $g\colon \Sigma\to G$. This is obvious for the condition $D_A\varphi\to 0$, since the covariant derivative transforms covariantly via the linear Higgs representation $\rho\colon G\to\text{GL}(V)$. Similarly the condition $\psi_\varphi\to 0$ is preserved since $\psi_\varphi$ also transforms covariantly.\footnote{To see this, recall that, in covariant notation, we have $D_\mu\varphi\to \rho(g)\cdot D_\mu\varphi$, so in particular $D_0\varphi\to \rho(g)\cdot D_0\varphi$. In the 3+1 formalism we work in the temporal gauge $A_0=0$ and replace $D_0\varphi=\partial_0\varphi-eA_0\varphi=\partial_0\varphi$ by $\psi_\varphi$.} This means that the conditions $\psi_\varphi\to 0$ and $D_A\varphi\to 0$ are automatically preserved, as 0 is mapped linearly to 0. The same goes for the condition $\varphi\to 0$, since $\varphi$ evidently transforms linearly under $\rho(g)$ with $g\in\mathcal{G}^I$. For pure Yang-Mills theory the group $\mathcal{G}^I$ consists of transformations that become constant asymptotically, and since the unbroken Higgs field boundary conditions do not impose any extra requirements on gauge transformations we find the same boundary-preserving (allowed) gauge group for Yang-Mills-Higgs theory.

\textit{The broken phase.} In the broken phase things are different. The asymptotic conditions are now $\psi_\varphi\to 0, D_A\varphi\to 0$ and $\varphi\to\text{min}$. That is, the Higgs field must become a covariantly constant minimum of the potential $V(\varphi)$, and its velocity must become zero. At first sight, this seems to still allow $\mathcal{G}^I$ to contain all asymptotically constant transformations. After all, a gauge transformation maps a minimum of $V(\varphi)$ to another minimum. However, this is wrong, for the same reason as for pure Yang-Mills theory. Allowing for gauge transformations which act at infinity gives rise to a nontrivial configuration space $\Tilde{Q}_\infty$ at infinity. After all, if we let $\varphi_\infty$ denote some covariantly constant minimum at infinity, then $\Tilde{Q}_\infty$ will consist at least of an orbit of $\varphi_\infty$ under the action of $\mathcal{G}^I$ at infinity, i.e. of the constant gauge transformations at infinity. But this means that the tangent space $T_{\varphi_\infty}\Tilde{Q}_\infty$ is far from being 0-dimensional. In fact, it has the dimension of $G$, since it is the tangent space to an orbit of the action of the constant gauge transformations, which orbit is isomorphic to $\mathfrak{g}$. But we cannot allow $T_{\varphi_\infty}\Tilde{Q}_\infty$ to contain nonzero vectors, since we required that the tangent vectors $\psi_\varphi\in T_\varphi\Tilde{Q}$ vanish at infinity! Like for pure Yang-Mills theory, the requirement that velocities vanish asymptotically makes the Higgs field non-dynamical at infinity and forces us to impose a Dirichlet boundary condition on the Higgs field itself, even though only the tangent vectors and curvatures appear in the instantaneous Lagrangian.

Thus we are forced to require a stricter asymptotic boundary condition on the Higgs field: we need that $\varphi\to \varphi_\infty$, where $\varphi_\infty$ now denotes some \textit{fixed}, covariantly constant minimum at infinity. This ensures that the configuration space at infinity is zero-dimensional, consisting only of $\varphi_\infty$. The tangent bundle at infinity is therefore also zero-dimensional, consisting only of $\psi_{\varphi_\infty}=0$, as required for finiteness of energy. But clearly this stricter asymptotic boundary condition breaks gauge invariance, in the sense that it is not preserved by gauge transformations which act non-trivially at infinity. Only gauge transformations that are the identity at infinity preserve $\varphi_\infty\neq 0$, so we find that the groups of boundary-preserving and redundant gauge symmetries are equal up to connected components, i.e. $\mathcal{G}^I=\mathcal{G}^\infty$, where $\mathcal{G}^\infty$ denotes the group of gauge transformations that are constant at infinity (but only its identity component $\mathcal{G}^\infty_0$ is generated by the Gauss law constraint).

In this way we conclude that the physical gauge group $\mathcal{G}^I/\mathcal{G}^\infty_0$ equals (several copies of) the global gauge group in the unbroken phase but is discrete (trivial in the Abelian case) in the broken phase. This conclusion results from the difference of what we call the ``vacuum'' in the two cases: either $\varphi=0$ or a minimum of the potential $V(\varphi)$. Gauge symmetry breaking in the Higgs mechanism must therefore be understood as an alteration in the vacuum itself, leading to different asymptotic boundary conditions that ensure what it means to have finite energy.

\section{Conclusion}

In this paper we have given a rigorous identification of (several copies of) the group of global gauge symmetries with the quotient of asymptotic symmetries
\begin{align*}
    \mathcal{G}_\text{Phys}=\mathcal{G}^I/\mathcal{G}^\infty_0
\end{align*}
in Yang-Mills theory on a three-dimensional Euclidean Cauchy surface. Here $\mathcal{G}^I$ denotes the group of allowed or boundary-preserving transformations and $\mathcal{G}^\infty_0$ the group of transformations that are trivial in the sense that they yield the Gauss law constraint defining the constraint surface, and must therefore be viewed as redundant. Global gauge symmetries thus correspond to the physical asymptotic symmetry group. There were two main points to this derivation, corresponding to obtaining $\mathcal{G}^I$ and $\mathcal{G}^\infty_0$ respectively.

First, we found that instantaneous spatial asymptotic boundary conditions on Yang-Mills fields that ensure existence of the instantaneous Lagrangian only directly lead to requirements on tangent vectors $\alpha_A\in T_AQ$ and on the curvatures $F(A)$ of the connections $A\in Q$. However, the fact that the electric field vanishes on the boundary makes the gauge field non-dynamical there. Thus one finds that the instantaneous state space decomposes into a disjoint union of ``superselection sectors'' which one cannot move between. Instead of treating these sectors simultaneously, we chose to work in one by imposing one specific Dirichlet boundary condition, resulting in an instantaneous state space which is the tangent bundle to the space of all gauge fields satisfying this Dirichlet boundary condition. In a future work \cite{borsboominstantaneous} we will consider what happens when one works with the full instantaneous state space of fields in the presence of a boundary on which the velocity of the field must vanish. We will show that this leads to a stratified structure, in which every dynamical sector is labeled by a gauge field configuration on the boundary. We expect this stratified structure to be preserved by the instantaneous Legendre transform. The results of this paper can then be viewed as the specific case in which one restricts to one such sector/stratum by implementing a specific Dirichlet boundary condition.

Choosing one Dirichlet boundary condition corresponds to imposing asymptotic fall-off behavior on the gauge fields $A\in Q$ themselves. We have shown that it is not enough to require that gauge fields become flat at infinity, since this would still allow for non-zero tangent vectors $\alpha_A$ at infinity, which would spoil the finiteness of energy. Intuitively, this means gauge transformations acting at infinity create infinite energy, even if the energy depends only on gauge-invariant quantities. To counter this, we need to require that $A$ approaches a \textit{fixed} flat connection at infinity. In the non-Abelian case we also chose this Dirichlet boundary condition to be invariant under the adjoint action of the structure group $G$ in order not to restrict the gauge group unnecessarily. The gauge transformations that leave this fixed connection at infinity invariant are then precisely the elements of $\mathcal{G}$ that are constant at infinity. Properly interpreted, this means that we consider the equivalent problem on a conformal compactification $\hat{\Sigma}$ of $\Sigma$, and require that gauge transformations be constant on $\partial\hat{\Sigma}$. This yields the group of boundary-preserving gauge symmetries $\mathcal{G}^I$.

Second, we explained that redundant gauge transformations in the Hamiltonian formulation of Yang-Mills theory must be understood as the infinitesimal localizable symmetries $\mathfrak{G}$. These give rise to the Gauss law constraint and must therefore be interpreted as unphysical if (an appropriate form of) determinism is to survive. All infinitesimal symmetries in $\text{Lie}(\mathcal{G}^I)$ are localizable, except for the global ones. Thus $\mathcal{G}^\infty_0$ indeed consists precisely of all gauge transformations that are the identity at infinity and lie in the identity component of $\mathcal{G}^I$. Again, properly interpreted this means we move to $\hat{\Sigma}$ from $\Sigma$ and require elements of $\mathfrak{G}$ to be zero on $\partial\hat{\Sigma}$, so that elements of $\mathcal{G}^\infty_0$ are the identity on $\partial\hat{\Sigma}$. The quotient $\mathcal{G}_\text{Phys}$ then consists of a copy of the global gauge group $G$ for every homotopy class in $\pi_3(G)$.

Subsequently, we applied these ideas to Yang-Mills-Higgs theory, where we derived that $\mathcal{G}^I$ equals the group of asymptotically constant gauge transformations only in the unbroken phase. In the broken phase one can only permit asymptotically trivial transformations, for otherwise the action of the gauge group at infinity would create non-zero velocities of the Higgs field, carrying infinite energy.

In another article \cite{borsboomdeharo} one of us considers the implications of this last result for our interpretation of gauge symmetry breaking. There it is argued that the Higgs mechanism must be understood as an instance of global gauge symmetry breaking, as has been proposed in the Abelian case for quantum electrodynamics \cite{morchioLocalizationSymmetries2007,depalmaNonperturbativeArgumentNonabelian2013,strocchiIntroductionNonPerturbativeFoundations2016,struyveGaugeInvariantAccounts2011a,borsboom2024spontaneous}. In future research it would also be of interest to extend our results to spacetimes with a nonzero cosmological constant and to better understand the relation of our work to asymptotic symmetries of Yang-Mills fields on the full boundary of spacetime (e.g. in celestial holography) \cite{rielloNullHamiltonianYang2025}, as well as to edge modes \cite{rielloEdgeModesEdge2021,carrozzaEdgeModesReference2022,ballDynamicalEdgeModes2024} and (quantum) reference frames \cite{delaHamette:2020dyi,Kabel:2024lzr,fewsterQuantumReferenceFrames2024}, and boundaries which are not asymptotic \cite{gomesUnifiedGeometricFramework2018,gomesUnifiedGeometricFramework2019,gomesGaugingBoundaryFieldspace2019,gomesQuasilocalDegreesFreedom2021,rielloHamiltonianGaugeTheory2024}. Additionally, the implications for quantum field theory should be investigated \cite{Rejzner:2020xid}.

\section*{Acknowledgements}
The authors want to sincerely thank Klaas Landsman and Manus Visser for providing detailed feedback on manuscripts, suggesting technical and conceptual improvements and for their general involvement and guidance. The authors are also grateful to Sebastian De Haro for his collaboration on this and related work. SB thanks Aldo Riello and Henrique Gomes for their comments and the insightful conversations in real life and over email. We also thank two anynonymous referees for their detailed comments.

\paragraph{Author contributions}
SB wrote the article and derived most results, HBP contributed with discussions, suggestions, feedback and supervision.

\paragraph{Funding information}
This work is supported by the Spinoza Grant of the Dutch Science Organization (NWO) awarded to N.P. (Klaas) Landsman.

\begin{appendix}
\numberwithin{equation}{section}

\end{appendix}





\bibliography{bibliography.bib}


\end{document}